\documentclass[modern]{aastex701ed}
\NewPageAfterKeywords
\begin{document}

\title{The Stellar Content of NGC~3603 Revisited: Is the IMF Top Heavy?}
\altaffiliation{This paper includes data gathered with the 6.5 meter Magellan Telescopes located at Las Campanas Observatory, Chile. It also uses observations made with the NASA/ESA {\it Hubble Space Telescope (HST)}, obtained at the Space Telescope Science Institute, which is operated by the Association of Universities for Research in Astronomy, Inc., under NASA contract NAS 5-26555. These observations were made under proposals GO-10602 (PI: Maiz Apellaniz) and GO-11626 (PI: Massey).}

\author[0000-0001-6563-7828]{Philip Massey}
\affiliation{Lowell Observatory, 1400 W Mars Hill Road, Flagstaff, AZ 86001, USA}
\affiliation{Department of Astronomy and Planetary Science, Northern Arizona University, Flagstaff, AZ, 86011-6010, USA}
\email{phil.massey@lowell.edu}

\author[0000-0003-2535-3091]{Nidia I. Morrell}
\affiliation{Las Campanas Observatory, Carnegie Observatories, Casilla 601, La Serena, Chile}
\email{nmorrell@carnegiescience.edu}

\author[0000-0002-5787-138X]{Kathryn F. Neugent}
\affiliation{Lowell Observatory, 1400 W Mars Hill Road, Flagstaff, AZ 86001, USA}
\email{kathrynneugent@gmail.com}

\author{Monica Herzog}
\altaffiliation{NSF REU student 2012.}
\affiliation{Lowell Observatory, 1400 W Mars Hill Road, Flagstaff, AZ 86001, USA}
\affiliation{Department of Astronomy and Planetary Science, Northern Arizona University, Flagstaff, AZ, 86011-6010, USA}
\email{monica.herzog12@gmail.com}

\author[0000-0001-7081-0082]{Maria R. Drout}
\affiliation{David A. Dunlap Department of Astronomy and Astrophysics, University of Toronto, 50 St. George Street, Toronto, Ontario, M5S 3H4, Canada}
\email{maria.drout@utoronto.ca}

\author{Caitlin O'Brien}
\altaffiliation{NSF REU student 2021.}
\affiliation{Lowell Observatory, 1400 W Mars Hill Road, Flagstaff, AZ 86001, USA}
\affiliation{Department of Astronomy and Planetary Science, Northern Arizona University, Flagstaff, AZ, 86011-6010, USA}
\email{cao87@cornell.edu}

\begin{abstract}

Studies of the resolved stellar populations of young massive clusters
have shown that the slope of the initial mass function appears to
be the same everywhere, with no dependence on stellar density or
metallicity.  At the same time, studies of integrated properties
of galaxies usually conclude that the IMF does vary, and must be
top-heavy in starburst regions.  In order to investigate this, we
have carried out a long-term project to characterize the massive
star content of NGC 3603, the nearest giant H\,{\sc ii} region, known
to have a rich population of massive stars.  We used both ground-based
and Hubble Space Telescope (HST) imaging to obtain photometry, and
employed Gaia to establish membership.  We obtained spectra of 128
stars using the Magellan 6.5~m telescope and HST, and combine these
data to produce a reddening map.  After analyzing the data in the
same way as we have for 25 other star-forming regions in the Milky
Way and the Magellanic Clouds, we find that the IMF slope of NGC~3603
is quite normal compared to other clusters, with $\Gamma=-0.9\pm0.1$.
If anything, there are fewer very high mass ($>65M_\odot$) stars 
than one would expect by extrapolation from lower masses.   This
slope is also indistinguishable from what several studies have shown
for R136 in the LMC, an even richer region.  We speculate that the
depreciation of the highest mass bins in NGC 3603, but not in R136,
may indicate that it is harder to form extremely massive stars at
the higher metallicity of the Milky Way compared to that of the
LMC.

\end{abstract}

\section{Introduction}

One of the outstanding puzzles of star formation concerns the apparent universality of the initial mass function (IMF). \citet{1955ApJ...121..161S}
showed that when stars are formed there are a decreasing number of stars with increasing mass, and that this was well represented by
a power law.  
If we describe the IMF as 
$\xi(m)\propto m^{\Gamma-1}$, then for stars more massive than 0.5$M_\odot$, $\Gamma=-1.3\pm0.5$ \citep{2007IAUS..241..109K}.  This $\Gamma$ is referred to as the ``slope of the IMF", and  its $\pm0.5$ scatter reflects the normal variation seen from one OB association or cluster to another, and does not seem to be correlated with metallicity or
star density \citep{MasseyGilmore,2011ASPC..440...29M}; it may simply represent the stochastical nature of sampling of a probability function \citep{2001MNRAS.322..231K,2002ASPC..285...86K}.\footnote{The canonical ``Salpeter" value is $-1.35$, although \citet{2011ASPC..440....3Z} has pointed out that if \citet{1955ApJ...121..161S} had used the modern value for the age of the Milky Way's disk, he would have derived $\Gamma=-1.05$ \citep{2019NatAs...3..482K}. We also note that  Salpeter's sample did not contain any O stars, although it is used as the standard for massive stars.}

This universality of the IMF is surprising. For instance, one would naively expect that the reduced effect of radiation pressure during the accretion phase of star formation would favor the formation of higher mass stars in lower metallicity environments such as the SMC compared to that of the Milky Way.  It must be that other factors dominate.  The persistence of the IMF despite such stellar feedback is one of the great challenges in understanding star formation (see, e.g., \citealt{2011ApJ...731...61E}). In fact, to some extent, star formation theorists are still struggling to understand why the stellar IMF is a power law at all; modern explanations focus on cloud fragmentation driven by supersonic turbulence and self-gravity, clump coalescence, and protostellar accretion (see, e.g., \citealt{2011MNRAS.410..788S} and references therein), but the relative importance of these processes is still uncertain.

However, an exception to this universality {\it may} exist where the star-formation rates are extremely high.  Some experts assert that the IMF is flatter (``top heavy") in such starburst regions.  This evidence comes primarily from unresolved stellar populations, and traces at least as far back as \citet{1977ApJ...217..928H}, who found that blue galaxies (and hence ones with lots of active star formation) must have IMF slopes flatter than Salpeter in order to match their broad-band colors and  H$\beta$ emission.  The notion was further ingrained in our collective consciousness as a result of the  \citet{1980ApJ...238...24R} study of the starburst galaxy M82.  (It was this study that coined the phrase ``top heavy" to indicate the relative absence of low- and intermediate mass stars.) However, subsequent work by the same group showed that it was not true \citep{1993ApJ...412...99R}.  The notion of a top-heavy IMF in starburst galaxies was further promoted by \citet{1990ASSL..160..125S} (despite the reversal on M82), although he notes that the results could be just as well explained if the upper mass limit was higher than 80-100$M_\odot$.  Similarly, \citet{2009ApJ...695..765M} invokes a top-heavy IMF to explain the correlation of flux ratio of H$\alpha$ to the far-UV to global properties in a sample of H\,{\sc i}-selected galaxies,  although systematic variations in the upper-mass limit would also match their models.   Not all studies of the integrated properties of starburst regions agree.  For instance,  \citet{2011ASPC..440..309L} finds no evidence of environmental effects on the initial mass function in starburst regions and distant star-forming galaxies using population synthesis models based on UV spectral lines.  A recent  study of the transient frequency in the lensed Spock arc by \citet{2025ApJ...988..178L} shows that the data are consistent with a Salpeter IMF, and not with a top-heavy IMF, contrary to the usual assumptions about star formation in the early universe.

Despite the results of studies of resolved stellar populations, the idea that starburst regions have top heavy IMFs has continued to gain traction, with the two nearest giant H\,{\sc ii} regions, NGC~3603 in the Milky Way and R136 in the LMC, often cited as examples where this is the case.  
In part this may be driven by the belief that from a theoretical perspective such regions ``should" have a flatter IMF.   Some may also take the canonical Salpeter $\Gamma=-1.35$ value as an absolute standard, not fully recognizing that slight variations are seen, as described above.  Finally, there is also the fact that researchers will derive slightly different temperatures and luminosities for the same stars.
In R136, \citet{MH98} find $\Gamma=-1.3\pm0.2$.  For a more extended region in 30 Dor, \citet{2018Sci...359...69S} derive $\Gamma=-0.90_{-0.37}^{+0.26}$, again, well within the normal variations seen for the IMF slope in young massive clusters, yet describe their result as showing an ``excess of massive stars." \citet{2020MNRAS.499.1918B} conducted a similar study in R136, and using more advanced modeling techniques, derive $\Gamma=-1.0\pm0.3$, which also agrees well with previous values.

The case for NGC~3603 is more confused.  \citet{2008ApJ...675.1319H} found an IMF slope of $\Gamma=-0.74^{+0.62}_{-0.47}$,
which they describe as top-heavy.  While the value is flatter than a Salpeter $\Gamma=-1.35$, it is still well within the range we see in less extreme regions, such as NGC~6611, for which  \citet{1993AJ....106.1906H} find $\Gamma=-0.7\pm0.2$.   The \citet{2008ApJ...675.1319H}  study was based purely
on near-IR photometry, and included only stars with masses less than $20M_\odot$.  Yet NGC~3603 is known to contain one of the largest collections of massive stars known in the Milky Way, with $\sim$50 known O-type stars, including a dozen of the hottest and most luminous stars known \citep{1995AJ....110.2235D, 2008AJ....135..878M}. The brightest star in NGC~3603 is designated A1, and is a double-line binary, with masses as high as any known in the Milky Way,
93$M_\odot$ and 70$M_\odot$ \citep{MasseyA1}.  Within the Local Group, only the R136 cluster may outstrip NGC~3603 in its massive star content (e.g., \citealt{1994ApJ...436..183M, 2014A&A...564A..40W}).

Given the large number of extremely massive stars known in NGC~3603, we thought potentially it really might have a top-heavy IMF.
Our earlier study \citep{2008AJ....135..878M} whetted our appetite, and we continued observing its brightest stars spectroscopically with the Magellan 6.5~m telescopes once that project had finished.  The central cluster is very dense,  and we were fortunate to have obtained time with the Hubble Space Telescope (HST) to obtain spectra of many of its most crowded stars. In this paper we use photometry from HST and ground-based imaging with newly obtained spectra to reanalyze the massive star content of the cluster.  In Section~\ref{Sec-obs} we describe our observations and reductions, including the use of Gaia data to identify members.   In Section~\ref{Sec-class} we present our spectral classifications and identify newly found binaries.  We give our analysis in Section~\ref{Sec-analysis}, presenting a color-magnitude diagram (CMD) and producing a reddening map.    In Section~\ref{Sec-HRD} we construct the H-R diagram after deriving physical properties from our new data, and compare the stars' locations to evolutionary tracks.    In Section~\ref{Sec-IMF} we derive the slope of the IMF.  A summary of our work, and a discussion of our results, can then
be found in Section~\ref{Sec-discuss}.

\section{Observations and Reductions}
\label{Sec-obs}

\subsection{Photometry}

\citet{2008AJ....135..878M} describe their photometry using archival {\it HST} images obtained with the High-Resolution Camera (HRC) of Advanced Camera for Surveys (ACS).  The data had been taken through the F435W and F550M filters as part of program 10602 (PI: Jes\'{u}s Ma\'{i}z Apell\'{a}niz),  and had consisted of four dithered exposures, totaling 8 s in each of the two filters.  \citet{2008AJ....135..878M} had performed aperture photometry on the images, correcting the values for charge transfer efficiency losses, and transformed the photometry
to the standard Johnson B and V system rather than remaining in the ``native" ACS/HRC system.   Complete details are given in Section 2.3 of \citet{2008AJ....135..878M}.  We repeated their procedure here, but used the point-spread-fitting program {\sc daophot} \citep{1987PASP...99..191S} as implemented in IRAF\footnote{NOIRLab IRAF is distributed by the Community Science and Data Center at NSF NOIRLab, which is managed by the Association of Universities for Research in Astronomy (AURA) under a cooperative agreement with the U.S. National Science Foundation.}\citep{1986SPIE..627..733T,1993ASPC...52..173T,2024arXiv240101982F} rather than relying on just aperture photometry.  For most stars in common with \citet{2008AJ....135..878M} this resulted in negligible differences ($<$0.01-0.02~mag) with the values presented in their Table 1, but it allowed us to resolve a few stars that were blended in their study (for instance, NGC~3603-33).   We also extended their photometry to slightly fainter stars adding 16 stars by hand. Their study was also directed primarily at stars for which they had spectroscopic information, while our goal here is broader, namely to help characterize the entire massive star population of the cluster.  There are 213 stars with final photometry from the ACS/HRC data. 

The ACS/HRC images covered only the small, crowded central region of NGC~3603, roughly 30\arcsec\ on a side.  To extend our coverage, we used data obtained with the 1.3 m telescope located on Cerro Tololo Interamerican Observatory taken with the ANDICAM imager built and operated by the SMARTS consortium.  The CCD was a Fairchild 447 with a pixel scale of 0\farcs371 pixel$^{-1}$ on the 1.3 m. The data were taken on (UT) 2008 Jan 22 in queue mode, and consisted of twelve exposures: three 80 s B, three 20 s B, three 40s V, and three 10s V.  The consortium delivered reduced, flat-fielded images that were 950 $\times$ 856 pixels, corresponding to 5\farcm9 (EW) by 5\farcm3 (NS), or 0.009 deg$^2$. The delivered image quality was 3.2-4.3 pixels (1\farcs2-1\farcs6) on the twelve images. We performed digital photometry through a 5-pixel (1\farcs9) radius aperture.   Relative zero-point corrections were determined for the data taken in each filter,  and the photometry then averaged
in a filter-by-filter manner.  There were 757 stars in the final B catalog and 698 in the final V catalog, with 615 stars in common to the two. 

Although we would like to adjust the zero points of our ground-based data to those of the ACS/HRC photometry, there are two problems.  First, the ACS/HRC photometry covers only a very small range of colors since nearly all the  stars in the central cluster are O-type stars with similar reddening.  Thus, we needed high-quality photometry in common with the ground-based data to determine what color-terms, if any, to apply to our instrumental values.  The second problem is that only one of the stars in the central cluster covered by the ACS/HRC images, Sh 50,  is sufficiently isolated to make a reasonable comparison with our ground-based imaging.  Thus, to set the zero point for the ANDICAM data we used the photoelectric photometry of \citet{1978A&A....63..275V}.  These measurements were carried out for 23 of the brighter, least crowded stars identified by \citet{1965MNRAS.129..237S} in the NGC~3603 field, and were obtained using a system that is a much closer match to the original UBV system than any CCD observations are likely to be.  \citet{1978A&A....63..275V} warns that the $B$ band photometry in particular might be contaminated by nebula emission, and so we restricted our determination of the zero-points to the brightest stars in his sample.  

The resulting transformation equations showed no color term in $V$ and a slight color-term in B, $\sim$$0.061\times (B-V)$, fairly typical in our experience.  When we applied these transformation to the ANDICAM data we were pleased to find both the ACS and ANDICAM  photometry of Sh 50 yielded the same value for $V$, 14.72 ($\pm0.03$ for ACS, $\pm0.01$ for ANDICAM), and that the agreement at $B-V$ was reasonable, with values of 1.14$\pm0.05$ and 1.17$\pm0.01$, respectively.  

Finally, we eliminated all stars from the ANDICAM photometry that were in common with the ACS.  This left 551 stars (from 609) added from the ANDICAM photometry to the final catalog, which contains 764 stars in total.  We list these stars in Table~\ref{tab:photspec}, where we indicate whether the photometry comes from the ACS field or from the extended ANDICAM region.    In identifying the stars, our preference was not to give new designations to those stars already in the literature.  Thus we generally retain
the names provided by \citet{1965MNRAS.129..237S, 1986AA...167L..15H,1994ApJ...436..183M}, and \citet{2008AJ....135..878M},
particularly in the case of stars with prior spectroscopy.   For the remaining stars, we have added our own numbering, an extension of the numbers used in \citet{2008AJ....135..878M}. Stars with numbers of 2000 and above are all stars without spectroscopy, and the additional numbers in order of decreasing brightness.

\begin{deluxetable}{l c c c c c c c c l c l}
\tabletypesize{\scriptsize}
\tablecaption{\label{tab:photspec}Photometry and Spectroscopy of Stars Seen Towards NGC~3603}
\tablewidth{0pt}
\tablehead{
\colhead{ID\tablenotemark{a}}
&\colhead{RA$_{\rm J2000}$}
&\colhead{Dec$_{\rm J2000}$}
&\colhead{$V$}
&\colhead{$V_{\rm err}$}
&\colhead{$B-V$}
&\colhead{$B-V_{\rm err}$}
&\colhead{Field\tablenotemark{b}}
&\colhead{Gaia Mem.}
&\multicolumn{2}{c}{Spectroscopy} 
&\colhead{Comment} \\ \cline{10-11}
&&&&&&&&\colhead{Prob.}&\colhead{Type}
&\colhead{Ref.\tablenotemark{c}}
}
\startdata
2000&11:14:46.886&-61:16:00.63&10.78& 0.00& 0.23& 0.00&2& 0.000&            & &HD306201\\
Sh3 &11:15:23.852&-61:15:01.34&11.06& 0.00& 0.21& 0.00&2& 0.000&A0V         &1&HD306199\\
A1  &11:15:07.305&-61:15:38.43&11.18& 0.02& 1.03& 0.03&1& RUWE &O3If*/WN6   &4&        \\
2001&11:14:47.746&-61:13:17.40&11.24& 0.00& 0.25& 0.00&2& 0.000&            & &HD306198\\
B   &11:15:07.411&-61:15:38.58&11.33& 0.02& 1.01& 0.03&1& RUWE &O3If*/WN6   &4&        \\
Sh14&11:15:26.771&-61:17:48.40&11.65& 0.00& 0.16& 0.00&2& 0.000&            & &        \\
C   &11:15:07.589&-61:15:38.01&11.90& 0.00& 1.05& 0.00&1& 0.438&O3If*/WN6   &4&        \\
2002&11:14:51.826&-61:12:49.51&12.10& 0.00& 1.19& 0.00&2& 0.000&            & &        \\
Sh9 &11:15:15.157&-61:17:34.90&12.17& 0.00& 0.58& 0.00&2& 0.000&            & &        \\
Sh25&11:15:07.640&-61:15:17.53&12.33& 0.00& 1.50& 0.00&2& 1.000&B1Iab       &3&        \\
2003&11:15:25.492&-61:18:03.31&12.37& 0.00& 0.18& 0.00&2& 0.000&            & &        \\
201 &11:14:44.159&-61:14:42.48&12.51& 0.00& 1.38& 0.00&2& 0.000&late        &4&        \\
A2  &11:15:07.313&-61:15:38.79&12.53& 0.03& 1.03& 0.03&1&NoGaia&O3V((f))    &2,4&      \\
\enddata
\tablecomments{Table 1 is published in its entirety in machine-readable format.
      A portion is shown here for guidance regarding its form and content.}
\tablenotetext{a}{``Sh" designations are from \citet{1965MNRAS.129..237S}.  Double digit designations are from
\citet{1994ApJ...436..183M}.  Letter designations came originally from \citet{1928BAN.....4..261V} who visually identified 
several components in the central ``nebulous star" CPD  $-60^\circ$2732, which were and further resolved by speckle \citet{1986AA...167L..15H} and subsequently {\it HST} imaging \citep{1994ApJ...436..183M,2008AJ....135..878M}. 
All other designations (i.e., three or four digit numbers)  are either from \citet{2008AJ....135..878M} or new here.
}
\tablenotetext{b}{Field: 1=ACS, 2=ANDICAM.}
\tablenotetext{c}{References for spectral types:  
1:  \citet{1949AnHar.112....1C};
2: \citet{1995AJ....110.2235D}; 
3:  \citet{2008AJ....135..878M}; 
4: This paper (some of which were reclassified from previous sources) based on newly obtained spectra.
5: \citet{2013MNRAS.435L..73R} based on NIR spectra, and here based on newly obtained optical spectra.
}
\end{deluxetable}

We confirmed that the ACS and ANDICAM photometry went equally deep.  Both have a drop off in their numbers at $V$=18-18.5, and $B$=19-19.5,
several magnitudes deeper than we need, as we will show in Section~\ref{Sec-tracks}.

During our work, we confirmed a significant shift from the ACS coordinate system to that of the standard ICRS system.  We had calibrated our ground-based images using the {\sc astrometry.net} software \cite{Lang}.  The offsets were 
$\alpha_{ACS}-\alpha_{ICRS}=+0.180\rm{s}$ and
$\delta_{ACS}-\delta_{ICRS}=-0.21\arcsec$.    After applying this, we found that the coordinates derived from the ACS agreed with Gaia positions to roughly $\pm$0\farcs05.
The ANDICAM coordinates agreed with Gaia positions to roughly $\pm$0\farcs2.

\begin{figure}
\epsscale{0.75}
\plotone{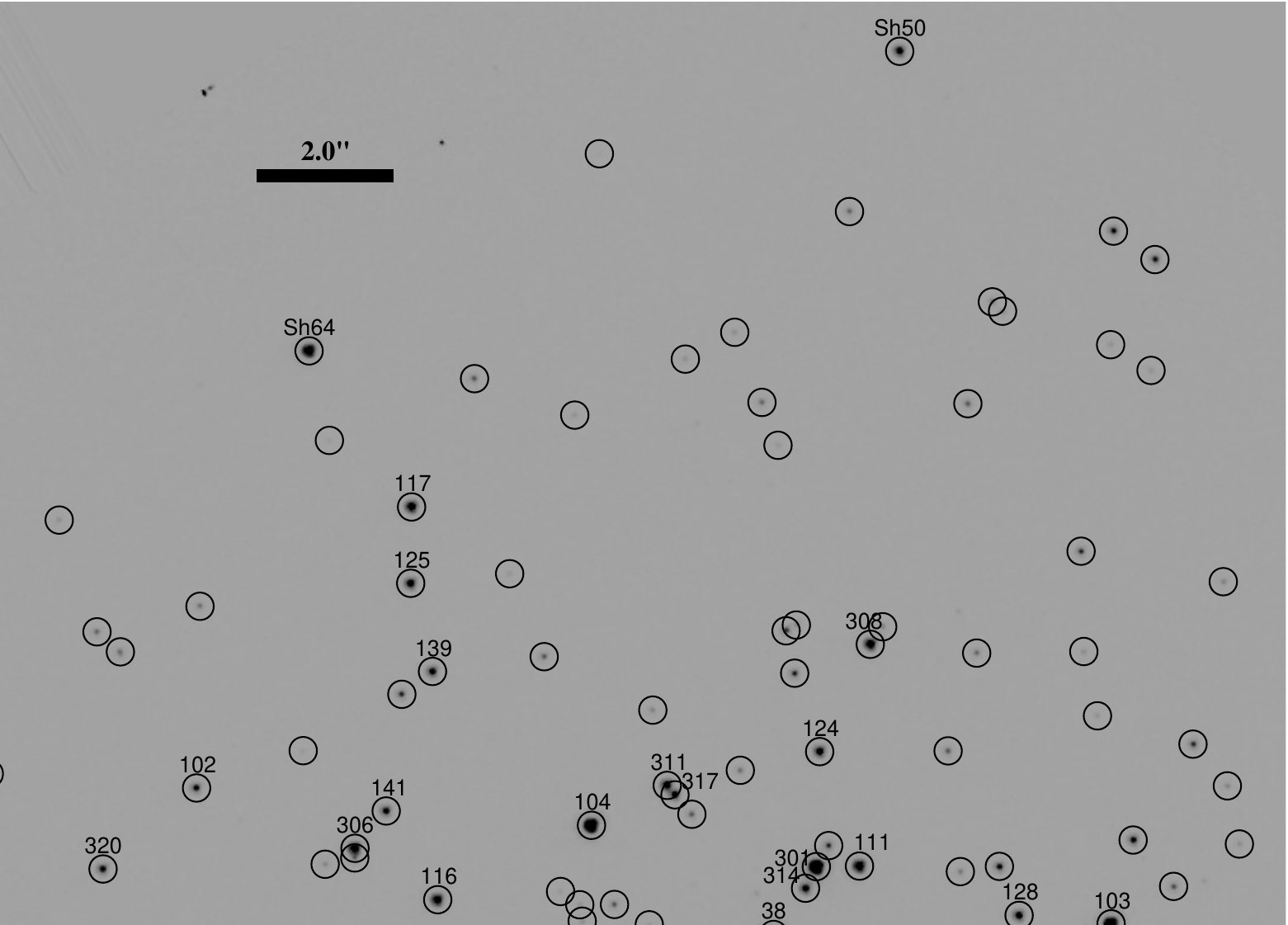}
\plotone{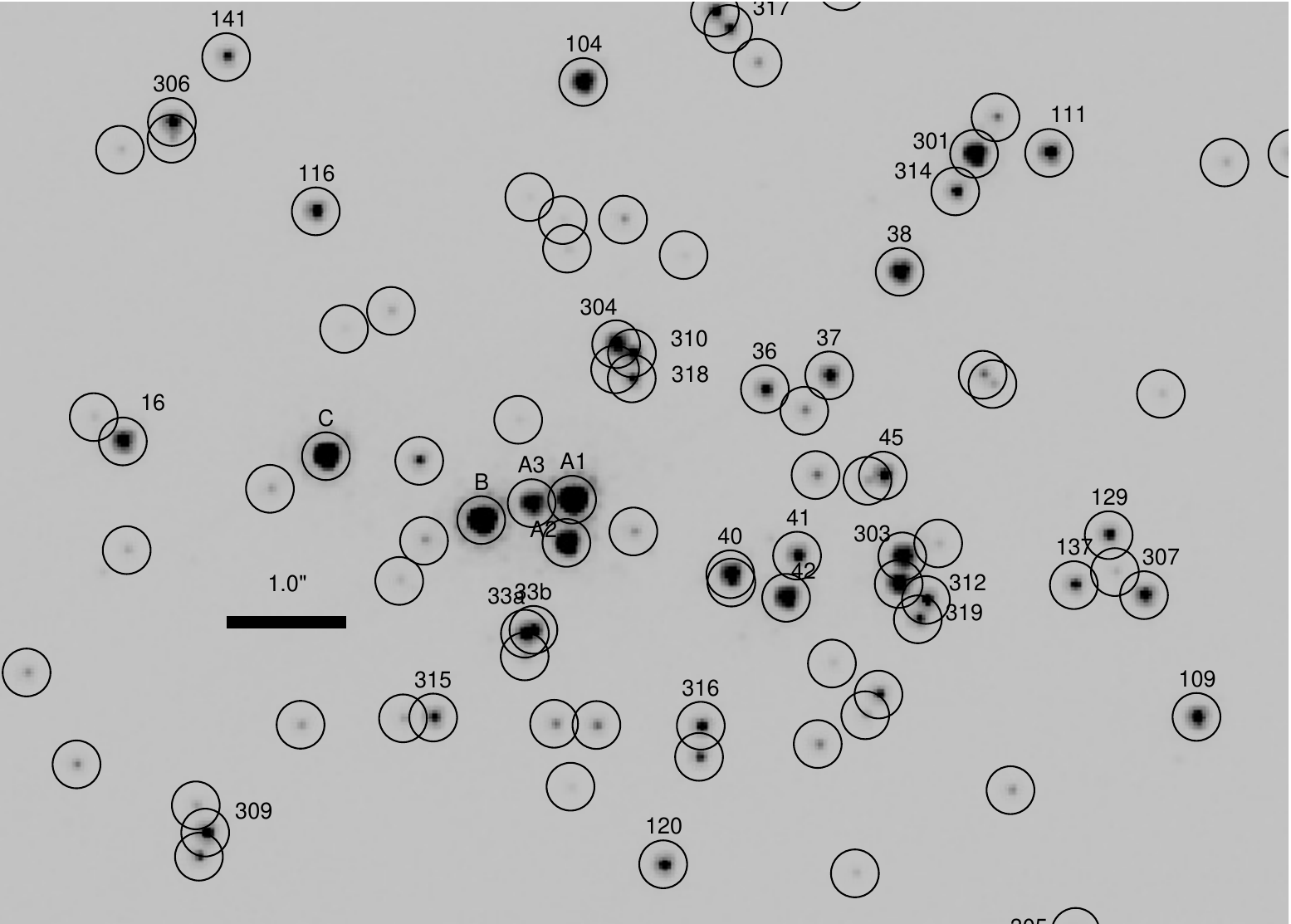}
\plotone{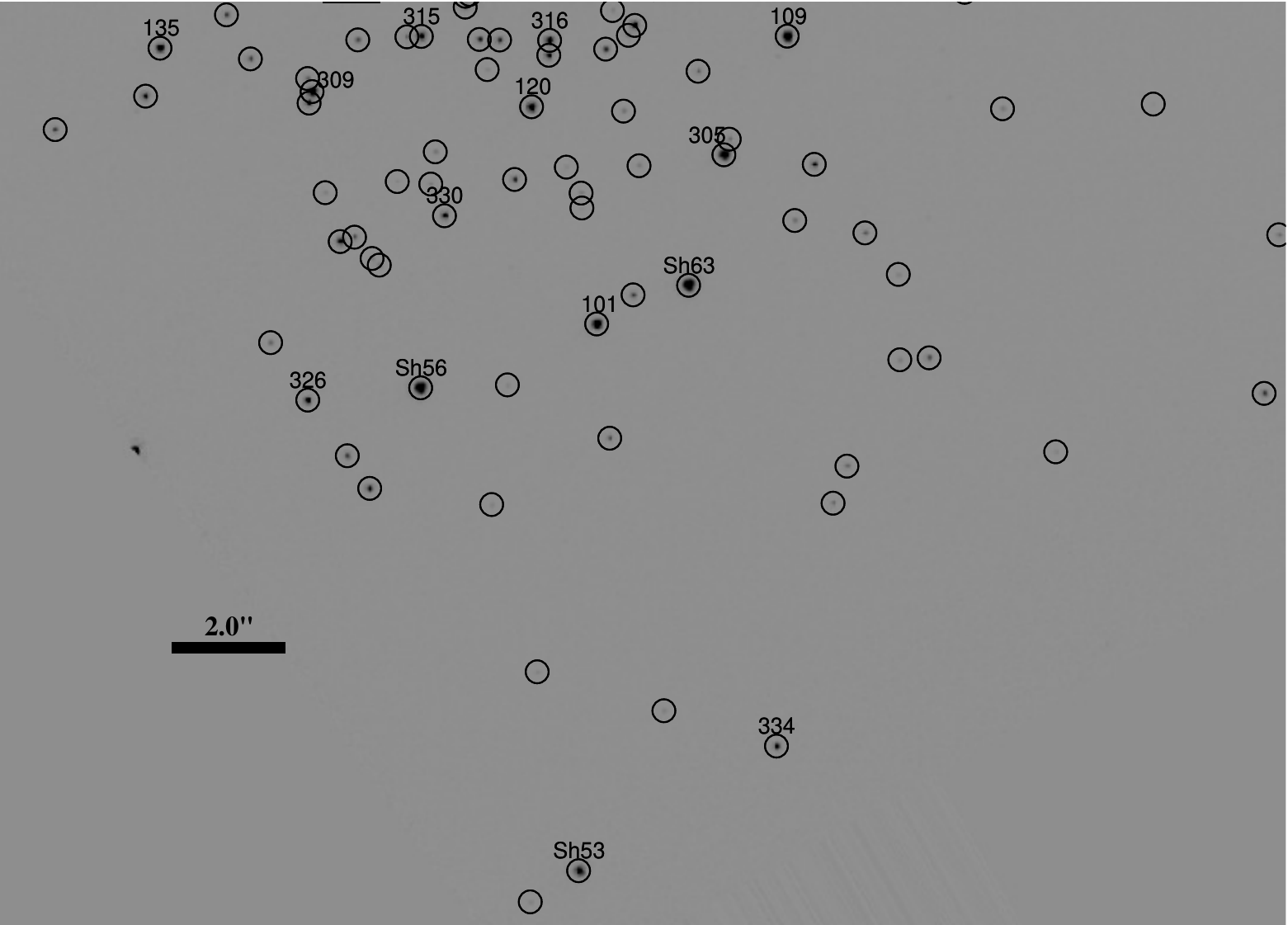}
\caption{\label{fig:ACS}  Program stars in the ACS field.  Stars with photometry on the ACS fields are encircled; those that also have spectral types are identified with their designations from Table~\ref{tab:photspec}.  The top panel is the northern portion of the ACS image; the bottom panel shows the southern portion. Each is roughly 16\arcsec\ on a side.
The middle panel shows the crowded central region at
an expanded scale, and covers roughly 6\farcs5 on a side.   The circles all have diameters of 0\farcs2. The ACS image has been rotated such that north is at the top and east is to the left.}
\end{figure}

\begin{figure}
\epsscale{0.90}
\plotone{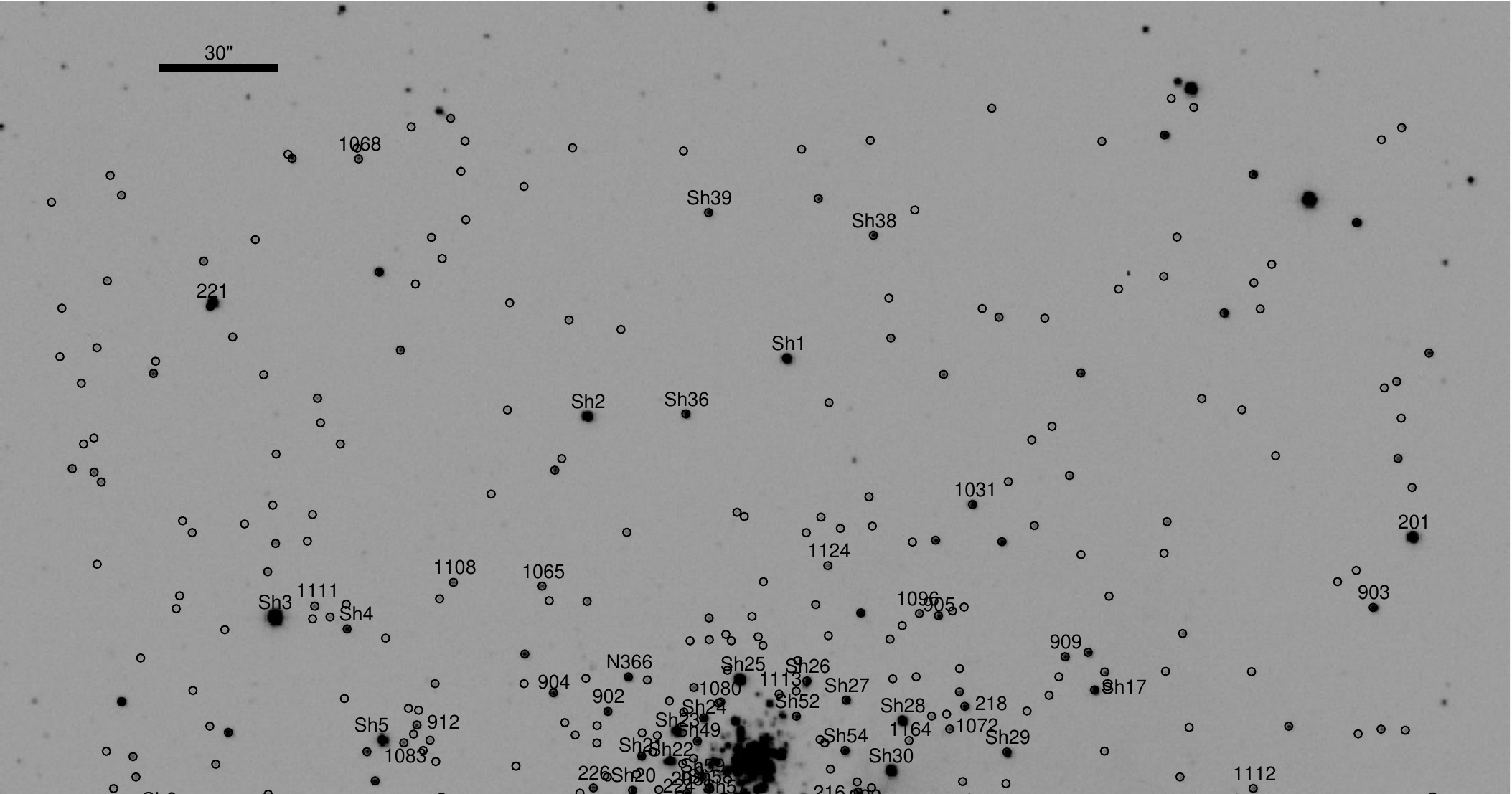}
\plotone{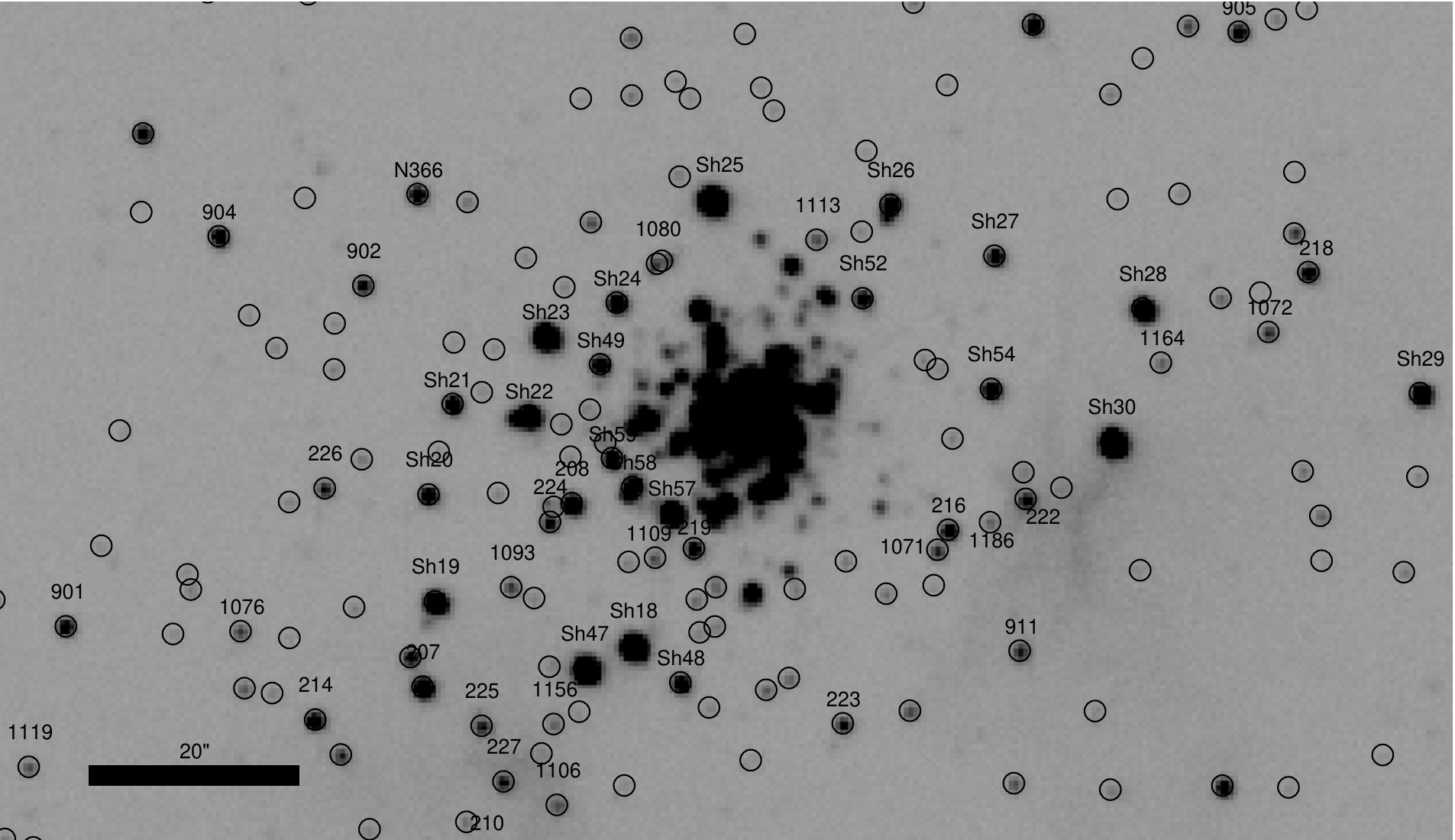}
\plotone{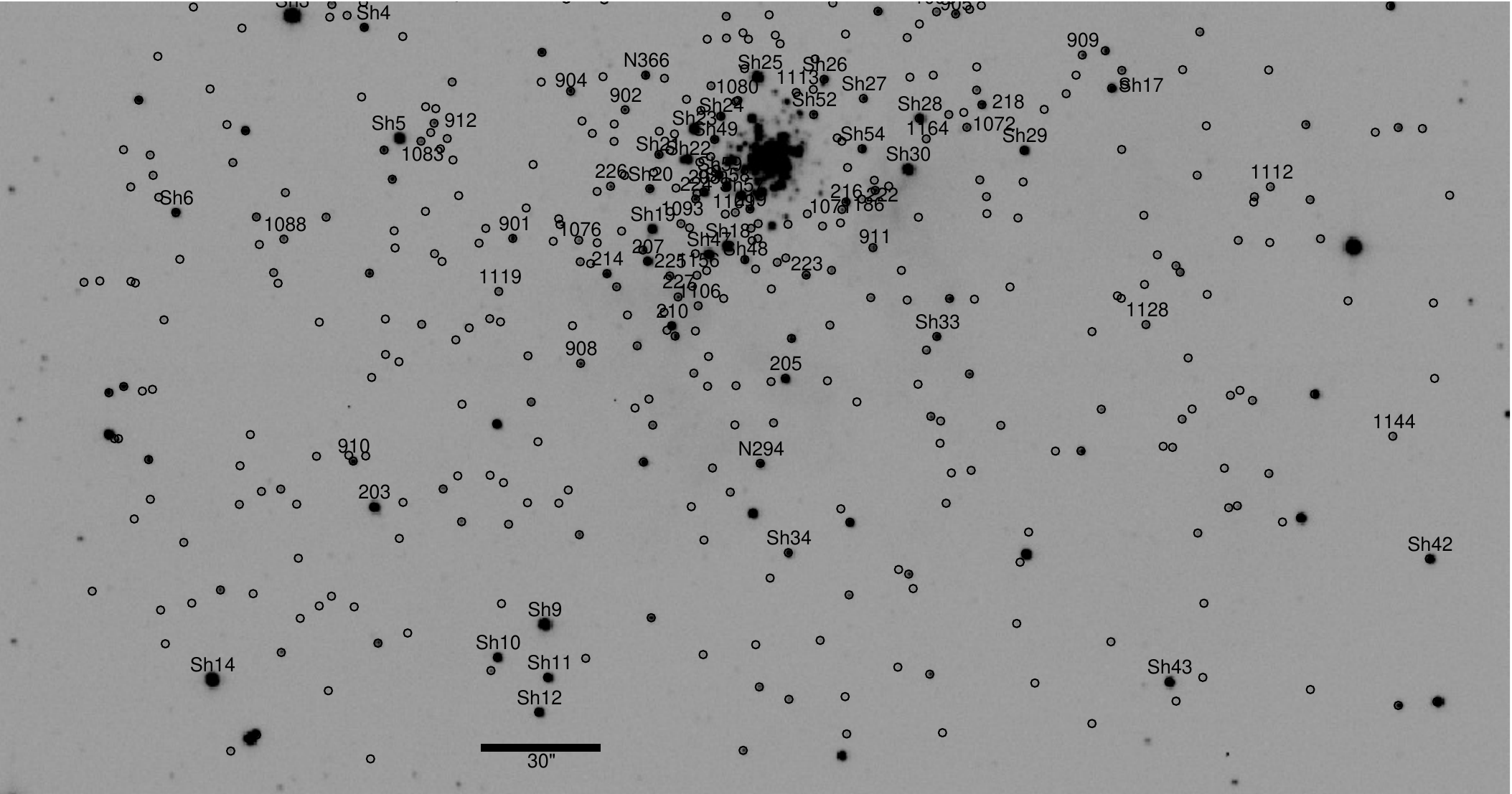}
\caption{\label{fig:ANDICAM}  Program stars in the wider ANDICAM field.  Stars whose photometry came from the ANDICAM data are encircled; those that also have spectral types are identified with their designations from Table~\ref{tab:photspec}.  The top panel is the northern portion of the ANDICAM field; the bottom panel shows the southern portion. Each is roughly 6\farcm3 $\times$ 3\farcm3 on a side.
The middle panel shows the more crowded region at
an expanded scale, and covers roughly 2\farcm3 $\times$1\farcm4 on a side.   The circles all have diameters of 1\farcs0. The image used here is part of a V-band exposure obtained with the Swope 1-m.}
\end{figure}

Stars with photometry are identified in Figures~\ref{fig:ACS} (ACS) and \ref{fig:ANDICAM} (ANDICAM).

\subsection{Membership Based On Gaia Proper Motions}
\label{Sec-gaia}
We next utilize astrometric data from Data Release 3 (DR3, \citealt{DR3, gaiadr3}) of the Gaia satellite \citep{gaia} to help determine membership.  Of our 764 stars, we have 
 identification of 612 sources (80.1\%) in the DR3 which also contain parallax and proper motion data.  Of these,
only 85 out of 213 ACS stars, or 40\%, had Gaia data, as might be expected due to crowding.   The ACS scale is 0\farcs05 pixel$^{-1}$, with a spatial resolution $<$0\farcs115 for well-dithered exposures.\footnote{https://hst-docs.stsci.edu/acsihb} Although the ultimate goal for Gaia is similar resolution,  currently the resolution for a single scan is 0\farcs23-0\farcs7, depending upon the orientation.\footnote{https://www.cosmos.esa.int/web/gaia/science-performance}

We identified 29 isolated stars for which we have OB-type spectra, and use these to define the average proper motions in 
right ascension and declination (pmRA and pmDEC), along with the intrinsic dispersion in these quantities, as well as the average parallax. 
We then performed a  covariance analysis in order to assign normalized probabilities for membership.
 Although Gaia parallaxes are notoriously unreliable beyond 2~kpc, we expect non-members to be foreground and not background stars, given the cluster's distance of 7.6~kpc  \citep{2008AJ....135..878M} and the fact we are working with luminous OB stars.  The mean proper motions derived from those 29 stars  are pmRA=-5.67 mas yr$^{-1}$ and pmDEC=+2.07 mas yr$^{-1}$, with an intrinsic dispersion of about 0.08 mas yr$^{-1}$ in RA and 0.06 mas yr$^{-1}$ in Dec.
The mean parallax is 0.1323 mas.  The results are about as expected: the analysis confirms that 80\% of the stars  in the ACS field with Gaia data are probable members, while only about 49\% of the stars with Gaia data in the extended ANDICAM field are likely members. 

The re-normalized unit weight error (RUWE) is a quantity provided with the DR3 data for every star; a value less than 1.4 is taken as an indication that the astrometric solution was well-behaved (see \citealt{2021A&A...649A...2L} as well as the unpublished review by L.\ Pearce 2021\footnote{http://www.loganpearcescience.com/research/RUWE\_as\_an\_indicator\_of\_multiplicity.pdf}). We found that many of the stars in the crowded ACS field had values of RUWE considerably higher than this cutoff.  For instance, A1, the brightest member of NGC3601, located deep in the core (see the middle panel of Figure~\ref{fig:ACS}), has a RUWE value of 13.6.  
However, if we used 1.4 as our cutoff, the number of stars for which we could use Gaia data as indicative of membership would drastically decrease, from 612 to 479.  We note that issues with the astrometric solutions are reflected in the uncertainties assigned to the proper motions; i.e., a star with a large RUWE value will also have large uncertainties in its astrometric properties, and thus our  covariance analysis probability assignment should be valid, in general.  The exceptions are stars with very large RUWE values, such as A1.  In examining our initial membership probabilities, we found that the handful of spectroscopically confirmed members with low membership probabilities mostly had RUWE values $>$2.6.  Thus we have decided, somewhat arbitrarily, to ignore the Gaia data for stars with  RUWE values of 2.5 or greater, and so note this in Table~\ref{tab:photspec} with the notation ``RUWE."

Two stars were initially missed in our photometry, NGC~3603-215 and -220.  Both are crowded: 220 is just to the east of the much brighter star Sh22,
and 215 is just slightly SE of Sh58. Both have Gaia data suggesting they are members, but their photometry is compromised by their close companions.  We have kept them since both have carefully obtained spectra; each are of type O7~V.  We exclude them from the CMDs and
the analysis of color excesses in the following sections but retain them in the HRD. 

\subsection{Spectroscopy}

\citet{1995AJ....110.2235D} reported spectral types for 12 stars from data obtained with the Faint Object Spectrograph
on-board the Hubble Space Telescope (HST).  Their study primarily found O3-type stars, which are among the hottest and most luminous stars known, similar to what
was found several years later in the R136 cluster by \citet{MH98}.  The \citet{1995AJ....110.2235D} work 
piqued our interest in the cluster, and we began our own spectroscopic
study.  We reported spectral types for an additional 16 stars in \citet{2008AJ....135..878M} using data from the Inamori Magellan Areal Camera and Spectrograph (IMACS, \citealt{IMACS}) on the Baade 6.5 m Magellan telescope at Las Campanas Observatory.  Those data were taken in long-slit mode using a 0\farcs7 wide slit with the long 
focal length (f/4) camera and the 600 line mm$^{-1}$ grating, yielding a spectral resolution of 2.0~\AA\ and covering the wavelength region 3600-7000~\AA.  These were also mostly of very early O-type, and we continued our spectroscopic investigation through 2023.  We  used the same IMACS setup to obtain spectra of additional NGC~3603 stars, as well as employing IMACS with multi-slit plates.  Additional spectra were obtained using the Magellan Inamori Kyocera Echelle (MIKE, \citealt{MIKE}) on the Clay 6.5 m Magellan telescope, and the Magellan Echellette Spectrograph (MagE, \citealt{MagE}) on either the Clay or the Baade, depending upon the year.   We also obtained spectra of the most crowded stars using the Space Telescope Imaging Spectrograph (STIS, \citealt{1998PASP..110.1183W,2024stis.rept....5R}).  We summarize our data sets in Table~\ref{tab:runs}.

\begin{deluxetable}{l l c l l l c c }
\tabletypesize{\scriptsize}
\tablecaption{\label{tab:runs}Spectroscopy Runs}
\tablewidth{0pt}
\tablehead{
\colhead{UT Date}
&\colhead{Instrument}
&\colhead{Slit Width}
&\colhead{Telescope}
&\colhead{Disperser}
&\colhead{Resolution}
&\colhead{Coverage}
&\colhead{Seeing}
}
\startdata
2006 Apr12,15\tablenotemark{a}  &IMACS f/4 Longslit & 0\farcs7 & Baade &600/8.6  & $\Delta \lambda$ = 2.0\AA & 3600-7000\AA\tablenotemark{b} & 0\farcs6-1\farcs1   \\
2008 Apr 21-23                             &IMACS f/4 Longslit & 0\farcs7 & Baade &600/8.6  & $\Delta \lambda$ = 2.0\AA & 3600-7000\AA\tablenotemark{b} & 0\farcs5-1\farcs0 \\                                                                                                                     
2008 Jun 14-16                             &IMACS f/4 Longslit & 0\farcs7 & Baade &600/8.6  & $\Delta \lambda$ = 2.0\AA & 3600-7000\AA\tablenotemark{b} & 0\farcs6-0\farcs9 \\   
2009 Mar 10, 12                            &IMACS f/4 Longslit & 0\farcs7 & Baade &600/8.6  & $\Delta \lambda$ = 2.0\AA & 3600-7000\AA\tablenotemark{b} &  0\farcs5-1\farcs0\ \\      
2010 May 15, 16, Sep 5                    &STIS/CCD              &0\farcs2  & HST    &  G430M, G750M& $\Delta \lambda $ = 1.1, 1.3\AA& 4050-4845\AA, 6300-6860\AA  &  \nodata  \\
2010 Mar 22, Jul 10, Oct 30              &STIS/CCD              &0\farcs2  & HST    &  G430M& $\Delta \lambda $ = 1.1\AA& 4310-4590\AA  &  \nodata  \\
2012 Feb 12, 13                            &MagE                      &0\farcs7,1\farcs0 & Clay     &Fixed prism & R=4100                               &3200-9500\AA   &0\farcs5-0\farcs9          \\
2021 Apr 21                                   &MIKE                       &1\farcs0 & Clay     & Fixed R2.4 &R= 28,000                          &3800-6100\AA   &1\farcs5-2\farcs2 \\
2021 Apr 22, 23                             &MIKE                       &0\farcs7 & Clay     & Fixed R2.4 &R= 40,000                          &3800-6100\AA  &0\farcs8-1\farcs3 \\
2023 Feb 27                                  &MagE                       &1\farcs0  & Baade    & Fixed prism & R=4100                         &3200-9500\AA   &0\farcs6-1\farcs0 \\
2023 Feb 28                                  &IMACS f/4 Multislit   & 1\farcs0   & Baade    & 1200/17.5& $\Delta \lambda$ = 1.3\AA  & 3850-5150\AA   &0\farcs7-0\farcs8 \\
\enddata
\tablenotetext{a}{Included in \citet{2008AJ....135..878M} }
\tablenotetext{b}{Gaps in coverage occurred at 4410-4440\AA, 5200-5230\AA, and 6000-6030\AA.}
\end{deluxetable}

The biggest challenge facing spectroscopy of the NGC~3603 stars is crowding. The most crowded stars were observed as part of a Cycle 17 HST program using a 0\farcs2 wide slit.  For the ground-based data, care was taken to observe crowded stars only in the best seeing and with slits widths of 0\farcs7. Less crowded stars were observed under worse conditions.  Additional  details are given below:

(1) The IMACS long slit Magellan observations usually were taken with the slit oriented in such a way that two or more stars could be observed simultaneously.  The 0\farcs7 wide slit covers nearly 15\arcmin\ in length, providing much flexibility in our choice of targets.  
Since the instrument has an atmospheric dispersion corrector (ADC), this could be done with little loss of flux even at the blue end.
The detector consists of a mosaic of 8 2K by 4K CCDs arranged in 2 rows (i.e., 8K by 8K in total), resulting in wavelength gaps unless two grating tilts are used.  For the  long-slit observations (2006-2009), we instead were
simply careful in our setup to assure no critical lines fell into these gaps. Given the fine spatial sampling of the unbinned CCDs (0\farcs11) and a best focus of 5 unbinned spectral pixels with the 0\farcs7 wide slit, we operated the detector in a 2x2 binned mode.   Flat field observations were obtained in the afternoon for removing pixel-to-pixel variations, and
a HeNeAr comparison lamp was obtained  after each observation for wavelength calibration. 

(2) STIS/CCD HST observations were made under program GO-11626  (PI: Massey), utilizing 28 orbits during 6 visits in Cycle 17,  with the explicit purpose of addressing the high-mass end of the mass function.  The goal was to obtain spectral classification of the stars that were too crowded to observe from the ground in the core of the cluster.  The 11 brightest stars were observed with four grating settings: G430M/4194, G430M/4451, G430M/4706, and G750M/6581.  The three G430M grating settings provided continuous coverage from 4050\AA\ to 4845\AA, a region which contains the spectral lines most critical for classification, while the G750M observation covered the H$\alpha$ line. At each grating setting, the star was dithered to five positions along the 0\farcs2$\times$52\arcsec\  slit, with steps of 0\farcs5 in order to reduce the effects of hot pixels and cosmic rays.  In addition, the longer exposures (typically $>$270 s) were also CR-SPLIT at each dither point.
An additional 25 stars were observed with just the critical G430M/4451 setting.  This region included H$\gamma$ and the
temperature sensitive He\,{\sc i} $\lambda$4471 and He\,{\sc ii} $\lambda 4542$ lines, upon which the classification of O-type stars is based \citep{1973ApJ...179..181C, WF, Sota}.
Observations were all made
by using offsets from Sh 23, carbon-rich O-type supergiant (OC9.7~Ia) with  V=12.7.    The standard pipeline reductions were used up through wavelength calibration and flat-fielding,  with the final extraction done using {\sc iraf} on the pipeline ``flt" files in order to utilize optimal extraction techniques (see, e.g., \citealt{MasseyHanson}). A small, $\pm$3 pixel aperture was used to extract each spectrum, with a modal background subtracted using  regions on either side of the star.  The extracted spectra were traced in the spatial direction.  For the 11 bright stars with 5 dither positions, the spectra were then combined using the ``avsigclip" algorithm; this resulted in vastly cleaned spectra.  The final spectra of the three blue regions were then combined and the spectrum classified.

(3) MagE Magellan observations were made in 2012. Crowded stars were observed with the 0\farcs7 slit when the seeing was excellent (0\farcs5-0\farcs6). When the seeing was not quite as good, we switched to the 1\arcsec\  slit and observed less crowded stars. Calibration and data reduction relied in a combination of {\sc mtools} routines (written by Jack Baldwin) and IRAF echelle tasks, as described in \citet{2012ApJ...748...96M}.

(4) MIKE Magellan observations were made during the first half of three nights in April 2021.  The seeing was particularly atrocious on the first night, requiring a wider slit (1\arcsec) than we preferred, but improved significantly for the next two nights, during which we used the 0\farcs7 slit.  MIKE had recently been outfitted with an ADC, but it became clear from the first few observations that it was misaligned, and it was not used during the run. Flat-field observations were obtained during the afternoons by the use of a diffuser, and a ThAr  exposure was obtained at one hour intervals for wavelength calibration. The data reduction  used the same combination of {\sc mtools} and {\sc iraf} echelle routines previously mentioned. 

(5) Both MagE and IMACS Magellan observations were made during February 2023.  As we shall see shortly, we had mostly found O-type stars, presumably of high masses.  Our goal for this run was to go fainter, reaching down to B-type dwarfs.  Our initial plan was to use both nights for IMACS observations with multislit plates. However, we learned upon our arrival  that there were intermittent electronic failures rendering the use of chips c1-c4 problematical.  Flat-field exposures through our masks during the afternoon, however, confirmed that the science targets all fell on chips c5-c8.  Unfortunately, the alignment stars, needed for adjusting the mask in x and y as well as rotation, were distributed throughout all 8 chips, and the alignment
routines required the software to be able to examine all 8 chips.  
Dr.\ Carlos Contreras was kind enough to modify these routines in real time, despite being off-shift.  We had difficulties with the alignments on the first night (due to user error on the part of PM) and switched to MagE to observe
stars that did not fall on a mask.  On the second night we observed all three masks, with the grating tilt adjusted between
exposures in order to fill in the gaps.  (Where a particular spectral line would fall depended upon the spatial location of the slit on the mask, and this process assured that no critical line would be lost.)  Each mask was observed for 3x1200s at each of the two grating tilts, with HeNeAr comparisons and flat-field exposures done on-sky for each tilt.
The data were reduced using the {\sc cosmos3} software 
with aid from Drs.\ Gus Oemler and Daniel Kelson, who helped us deal with the fact that half the array was missing from the data.  All together 33 spectra (including some repeats) were obtained using three multislit masks.

\section{Spectral Classification}
\label{Sec-class}
Stars were classified by referencing to the spectral atlases of \citet{WF}, \citet{WalbornO2}, and \citet{Sota}, with additional reference to \citet{2009ssc..book.....G} for the scant number of B-type stars and later.  For stars observed only with {\it HST}, with its limited wavelength coverage, we measured the ratios of the equivalent widths of the He\,{\sc i} $\lambda$4471 and He\,{\sc ii} $\lambda$4542 lines,
and used the quantitative classification criteria of \citet{1973ApJ...179..181C}, with luminosity classes assigned based on brightness.
The latter readily distinguishes O supergiants from dwarfs, but may confuse bright giants with the former, and faint giants with the latter. 
We list these spectral types in Table~\ref{tab:photspec}, and identify those stars for which we have spectroscopy by their designations in Figures~\ref{fig:ACS} and \ref{fig:ANDICAM}.

\begin{figure}
\epsscale{1.0}
\plotone{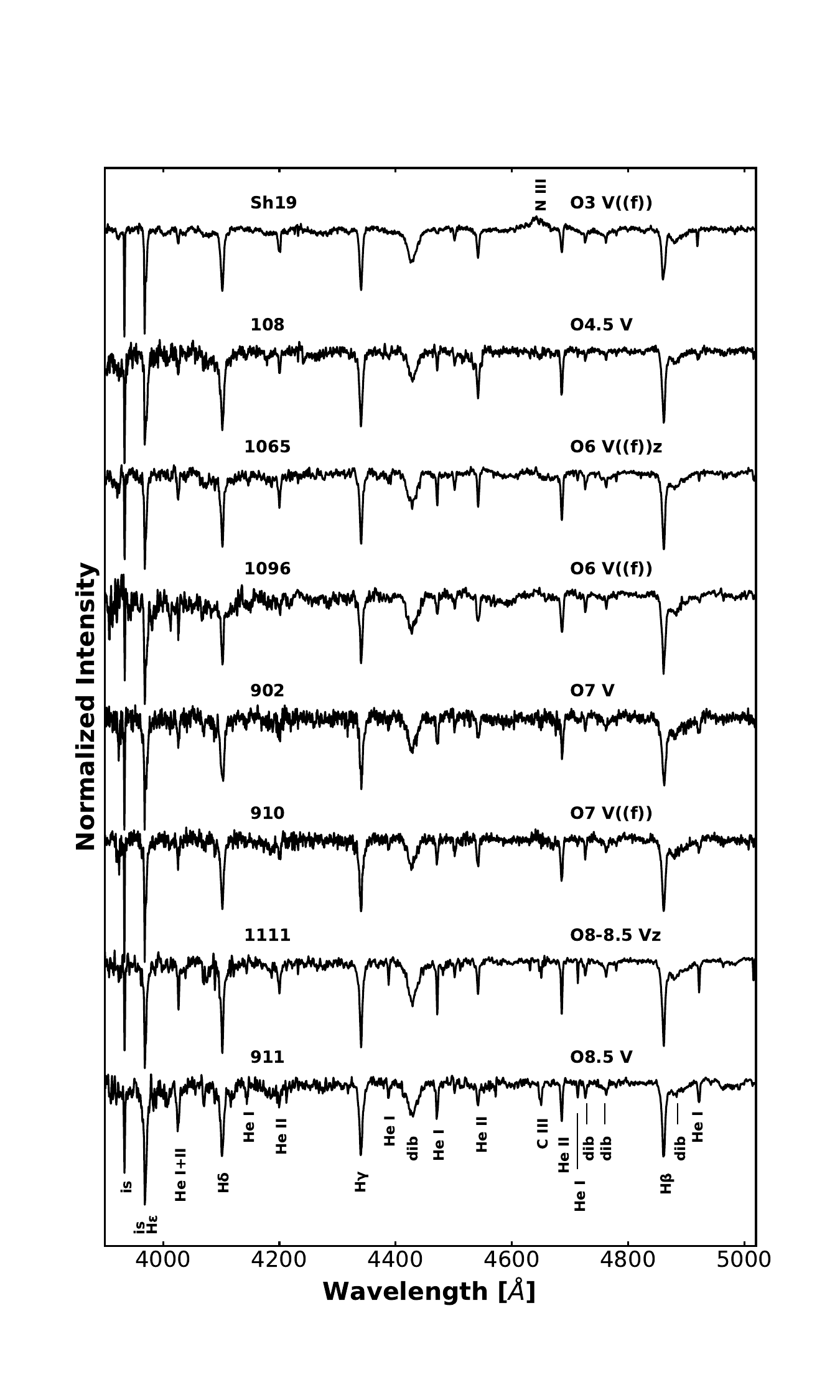}
\caption{\label{fig:spect1} Representative spectra for early- to mid O.  The spectra have all been normalized.  We label each with
its name and the adopted spectral types from Table~1.  The principal spectra features are identified at the bottom, with both
interstellar (is) lines and diffuse interstellar bands (dib) noted.}
\end{figure}

\begin{figure}
\epsscale{1.0}
\plotone{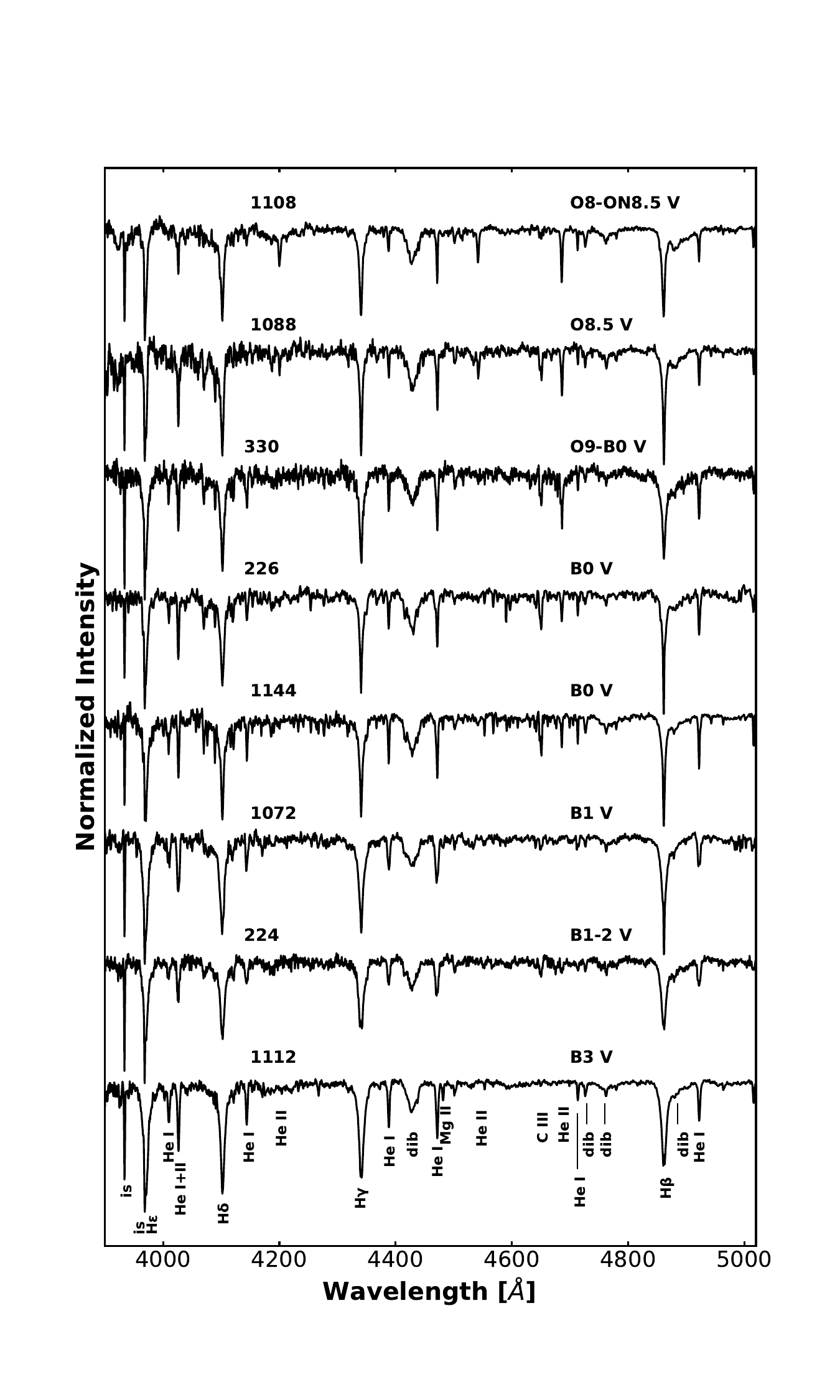}
\caption{\label{fig:spect2} Representative spectra for mid-O to early B.  The spectra have all been normalized.  We label each with
its name and the adopted spectral types from Table~1.  The principal spectra features are identified at the bottom, with both
interstellar (is) lines and diffuse interstellar bands (dib) noted.}
\end{figure}

In Figures~\ref{fig:spect1} and \ref{fig:spect2} we show representative examples of our spectra.  The data have been normalized,
and we have identified the principal spectral features. 

\subsection{O2-3If*/WN5-6 Stars and Other Stars from the Literature}
Certain classifications require more discussion.  The most luminous stars in young, highly populated clusters such as NGC~3603, R136 in the LMC, and NGC 604 in M33, have spectral characteristics that are intermediate between the hottest O-type supergiants (O2-3.5~If*) and mid-excitation WN-type Wolf-Rayet stars (WN5-6).  \citet{1982ApJ...254L..15W} introduced the notation O3~If*/WN6 to describe the star
Sk $-67^\circ$22 in the LMC.   Such stars are often referred to as ``slash" stars (e.g., \citealt{1989ASSL..157..297C}), or  characterized as ``Of-stars on steroids" \citep{MH98}.
These are unlike normal WN stars, in that they are
hydrogen-rich, possess absorption lines, and are much more luminous.  They are now understood to still be in the
hydrogen-burning phase \citep{deKoterR136,MH98} and show WN-like emission lines simply because they are so close to their
Eddington limit that their stellar winds are optically thick \citep{1998MNRAS.296..622C,2011A&A...535A..56G,2012ASPC..465..196C}.  
 \citet{2008AJ....135..878M} did not have new spectra of their own of NGC~3603-A1, B, or C, and simply quoted the ``WN6+abs" types given by \citet{1995AJ....110.2235D} based on optical {\it HST}/FOS spectroscopy.  Examining their spectra today we would
 call these all O3~If*/WN6, and we have revised their types accordingly in Table~\ref{tab:photspec}.  The star A1 is actually an O3~If*/WN5-6 pair of stars; its primary has the highest mass known from a direct Keplerian measurement \citep{1984ApJ...284..631M,2008MNRAS.389..806S,MasseyA1}.

 \citet{2012MNRAS.427L..65R,2013MNRAS.433..712R} reports discoveries of two
additional ``slash" stars in NGC~3603, each classified as O2 If*/WN6 on the basis of near-IR (NIR) spectroscopy.
The first of these \citet{2012MNRAS.427L..65R} designates as WR42e; it is the star we call 903.  \citet{2012MNRAS.427L..65R} suggests that it has been ejected from the cluster core since it is located in the outer region
of the cluster.  He notes that the ground-based proper motions are ambiguous on this issue.  Our optical spectrum shows the star
to be an O4~Ifc.  We can exclude the O2~If*/WN6 classification on a number of grounds.  First, He\,{\sc i} $\lambda 4471$ is clearly
present; its strength relative to that of He\,{\sc ii} $\lambda 4542$, along with the appearance of the rest of the spectrum is consistent with an O4 type.  Although He\,{\sc ii} $\lambda 4686$ is in emission, its equivalent width is -5~\AA, what we would expect of an   O-type supergiant, but not of a WN star.  Finally, although N\,{\sc iv} $\lambda 4058$ is in emission, its strength relative to N\,{\sc iii} $\lambda \lambda 4634,42$ is like that of other O4~If stars.   The ``c" is added simply to denote that the carbon lines are strong.

As for NGC~3603-903 having been ejected from the cluster's core: its Gaia DR3 proper motion is in excellent agreement with that of the cluster as a whole, i.e., pmRA=$-5.679\pm0.02$ mas yr$^{-1}$and $+1.908\pm0.02$ mas yr$^{-1}$, compared to that of the cluster quoted above, pmRA=$-5.67$ mas yr$^{-1}$ and pmDEC=$+2.07$ mas yr$^{-1}$, with intrinsic dispersions of 0.08 mas yr$^{-1}$ and 0.06 mas yr$^{-1}$, respectively.
We note that our study has found lots of other early-type stars this far from the cluster core; one may reasonably assume that they
formed there.

\begin{figure}
\epsscale{0.5}
\plotone{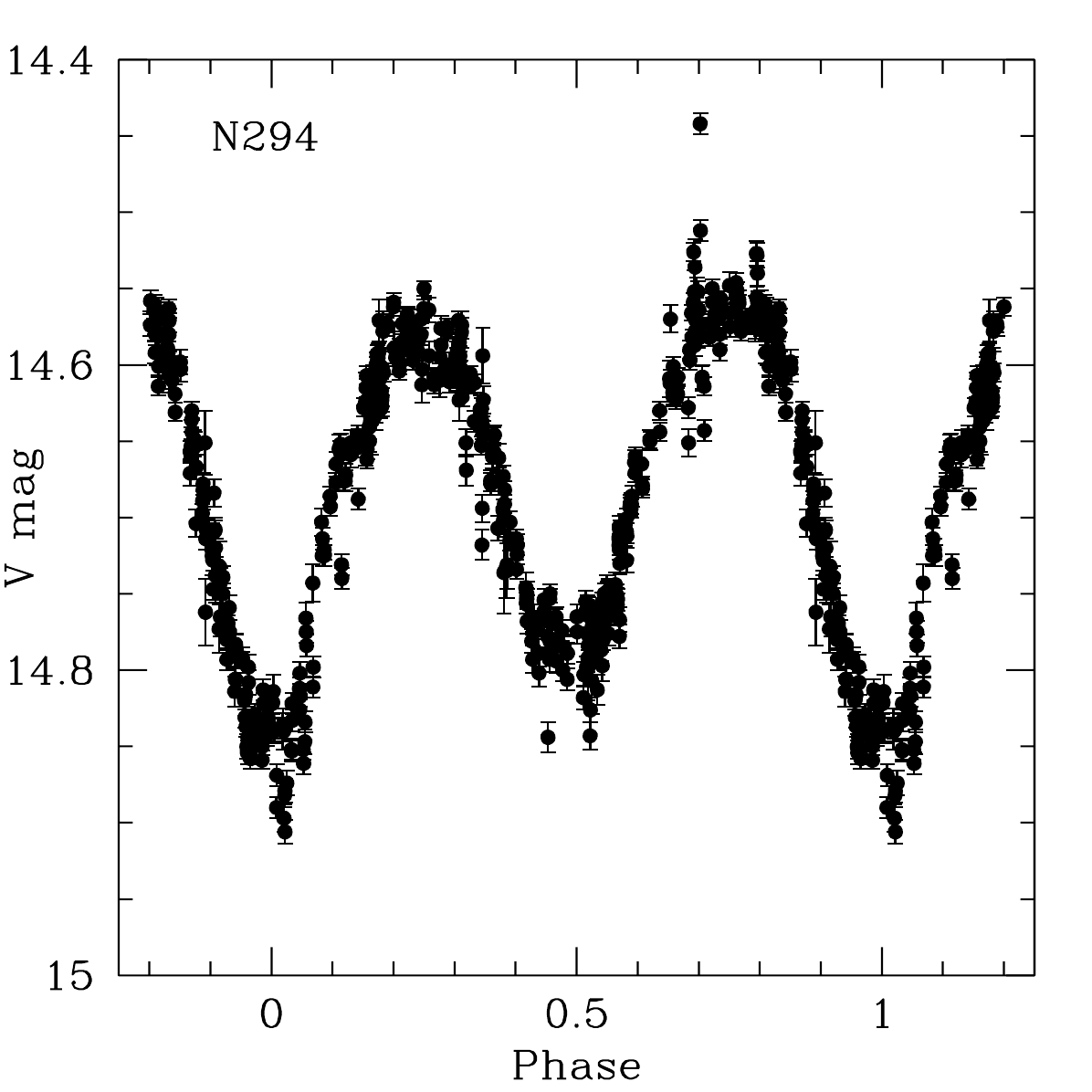}
\caption{\label{fig:N294lc} Light-curve of N294.  The data were collected in queue mode between January 2006 and June 2007 on the Yale 1-meter telescope
on Cerro-Tololo as part of the SMARTS program. 
The data have been phased with a period of 1.93557440 days and a phase zero-point T0 of 2453758.374.}
\end{figure}

The second of these slash stars, referred to as MTT 58 by \citet{2013MNRAS.433..712R},\footnote{The designation is from \citet{1989A&A...213...89M}.}  is the star we call N294.  From the near-IR spectrum, he arrives at an O2~If*/WN6 spectral type for the dominant star.  Its strong x-ray luminosity suggests it is a close binary,
and a careful inspection of their spectra led them to conclude the companion is an O3~If*.  Using SMARTS data obtained in the Yale 1-meter CTIO telescope in 2006 and 2007, we had already identified the star as a 1.94-day contact binary; its light-curve is shown in Figure~\ref{fig:N294lc}.   We continued to obtain Magellan optical spectra in order to determine an orbit solution for this massive pair.  We agree with the O2~If*/WN6+O3~If* classification of  \citet{2013MNRAS.433..712R}. 
We will describe the data and perform the orbital analysis in a subsequent paper, where we will discuss several other massive binaries in the NGC~3603 cluster.

A third massive star has been classified spectroscopically since \citet{2008AJ....135..878M}: also an x-ray source, the star is called MTT68 by \citet{1989A&A...213...89M},  and was classified as O2~If* by \citet{2013MNRAS.435L..73R}, also using NIR spectroscopy.  It is listed as
1031 in our Table~\ref{tab:photspec}. However,  our optical spectra suggests an O3~If spectral type.  We find that N\,{\sc iv} $\lambda 4058$ is stronger than N\,{sc iii} $\lambda 4634,42$, but not to the extent demanded for the O2 classification \citep{WalbornO2}.

\subsection{Binaries}
In the previous subsection we've discussed the binaries A1 (O3If*/WN6+O3If*/WN5, \citealt{MasseyA1}) and N294 (O2If*/WN6+O3If*, \citealt{2013MNRAS.433..712R}).  Here we list what we know about other identified binaries in the NGC~3603 cluster. 

{\it Sh56:} This star was classified as ``O3~III(f)+O?" by \citet{2008AJ....135..878M}.  We obtained a second spectrum of the star with IMACS on 2008 June 14 (see Table~\ref{tab:runs}).  Many of the He\, {\sc i} and He\, {\sc ii} lines are double.  Thus we confirm the fact that the star is composite, but cannot improve on the \citet{2008AJ....135..878M} classification.

{\it 207:}  We identified this star as an eclipsing binary as part of our SMARTS project, with a period of 1.07192 days.
 There are two equally
deep (0.4~mag) eclipses.   We classify the star here as ``O7V+O8V" based upon a series of ten spectra we obtained
between 2008 April 21 and 2010 February 23.    It is also listed in \citet{2023A&A...674A..16M} as a Gaia-detected eclipsing binary.  We believe this system deserves spectroscopic followup.

{\it 37:} The star was classified as ``O6.5+OB?" by \citet{1995AJ....110.2235D}.  We have no additional information about this system, and we
adopt their spectral type here. 

{\it Sh49:} We identified this as an eclipsing binary as part of our SMARTS project, with a period of 1.4858 days, and 0.2-0.4~mag eclipses.  The star was classified as ``O7.5~V" by \citet{2008AJ....135..878M}.  We have 10 spectra taken with IMACS and MagE between 2009 February 14 and 2012 February 13.  We see variations in the line profiles, and radial velocity variations of order 250 km s$^{-1}$ but no sign of double lines. 

{\it Sh4:} We classify this star as possibly composite (``O3.5IIIf + bin?") as the spectral energy distribution turns up at wavelengths above 5000\AA, and the strength of He\,{\sc i} $\lambda$5876 is much stronger than one would expect given its minimal presence at $\lambda$4471. Our guess
would be that the companion is an early B-type star. 

{\it N366:} This was also identified as an eclipsing binary as part of our SMARTS project with a period of 1.6113 days and shallow
eclipses (0.05-0.10~mag).  
We have 11 Magellan spectra (IMACS and
MagE) taken between 2008 April 22 and 2012 February 13.  We do not measure any significant radial velocity variations, nor do we detect any
double lines. 

{\it 223:} We classify this as B0.3III+early B, as some of the spectral lines (e.g., He\,{\sc i}$\lambda$4471) are double. Si\,{\sc iv} is not.
 
In addition, our modeling effort (Section~\ref{Sec-modeling}) failed to find acceptable fits for two stars:  Sh 18 and 208.  That leads us to suspect that each of them is composite. 

\section{Analysis}
\label{Sec-analysis}

\subsection{Color-magnitude diagram}
\label{Sec-CMD}
\begin{figure}
\epsscale{1.3}
\plotone{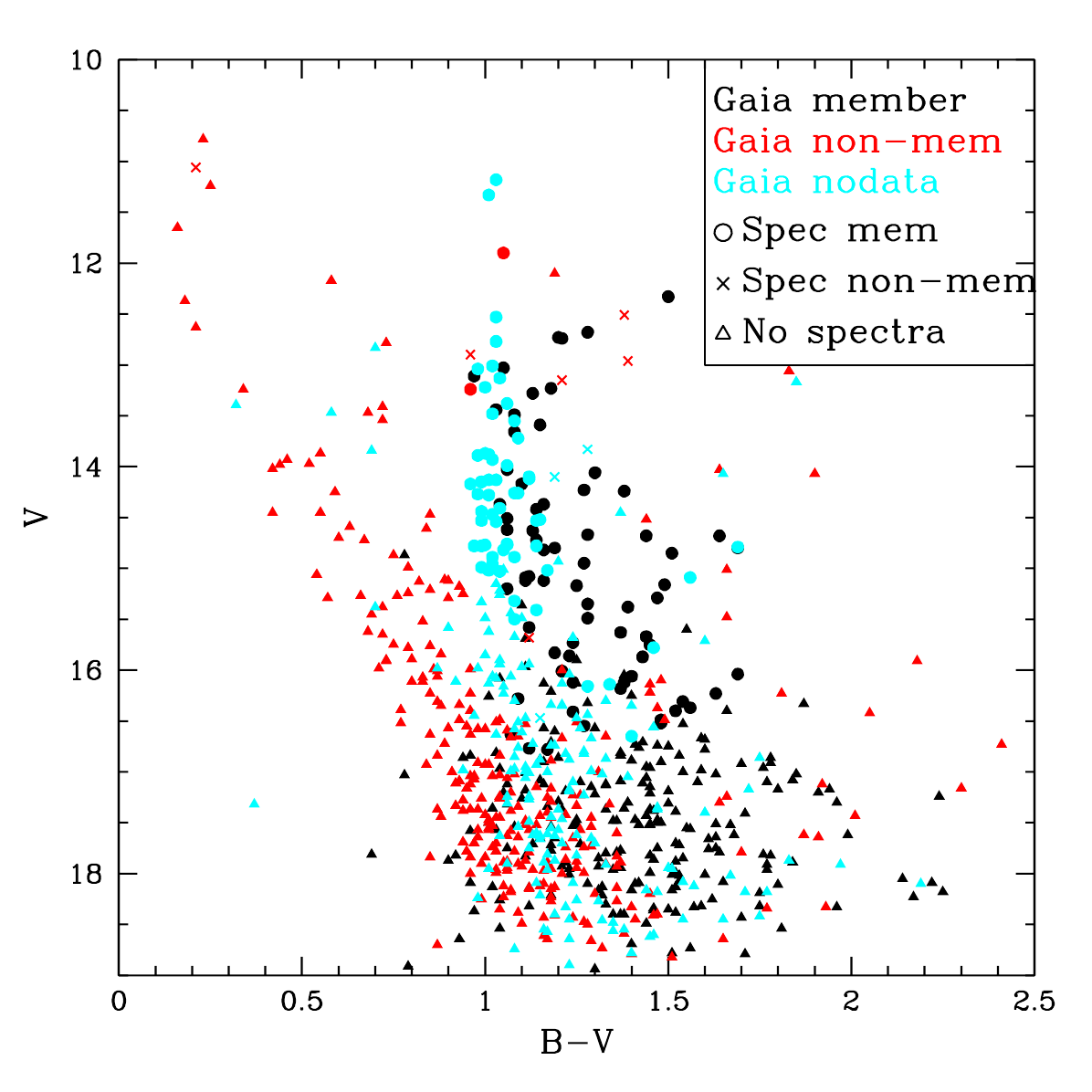}
\caption{\label{fig:cmd} Color-magnitude diagram for NGC~3603. As denoted in the legend, Gaia members are marked in black, Gaia non-members in red, and stars without reliable Gaia data marked in cyan. Circles denote stars which spectroscopy shows are members, x's denote stars confirmed to be non-members by spectroscopy, and triangles indicate stars without spectroscopy. There is a large swath of Gaia non-members extending from the brightest, bluest stars down to where cluster members are found at $V$$\sim$17, $B-V$$\sim$1.0. Cluster
members mostly occupy a color range of $0.9 <B-V<1.7$}
\end{figure}

In the previous sections we presented our photometry, determined the probability of membership based on Gaia proper motions, and described our spectroscopy.   In Figure~\ref{fig:cmd} we combine these data in a color-magnitude diagram (CMD) to get a more accurate assessment of the stellar content of the cluster, and the limitations of determining the IMF.

The red symbols denote stars whose Gaia data suggest they are not members (normalized probability $<$0.5 from Table~\ref{tab:photspec}).  Black is used for stars which Gaia data suggest are members, while cyan is used for stars lacking
adequate Gaia data (i.e., no proper motions or RUWE$\ge$2.5).   Spectroscopically confirmed members are shown as filled circles. 

We see that in general the cluster members  occupy a
color range from a $B-V\sim 0.9 $ to a $B-V\sim 1.7$, with foreground stars (red symbols) mostly occupying a diagonal band stretching from the brightest, bluest magnitudes ($V$$\sim$11, $B-V$$\sim$0.2) to where they merge with the cluster at $V$$\sim$17, $B-V$$\sim$1.0.

For the most part,  there is excellent agreement between the results of our spectroscopy and the Gaia proper motions once stars with high RUWE
values have been ignored.  The stars assigned to be members on the basis of Gaia data that also have spectroscopy indicating they are
members are shown by black circles; there are 71 such stars.  

Similarly, there are 6 stars, shown by red x's in Figure~\ref{fig:cmd}, where both the Gaia data and
spectroscopy demonstrate non-membership.  Four of these (Sh 1, Sh 2, NGC~3603-201, and -203) are bright, and fall into the color range $B-V=0.9$ to 1.4 where we expect to find members, but our spectroscopy shows these are late-type stars, in accord with their Gaia data.  A fifth example is Sh 3 (HDE 306199), the second-brightest star, shown by the red x in the upper left of the CMD.  This was identified as an A0 star in the Henry Draper Extension \citep{1949AnHar.112....1C}, and is certainly not a member.  A sixth star, NGC~3603-225, is confirmed as late-type in agreement with Gaia non-membership, despite its colors; it is much fainter ($V=15.68$) than the other five, near the limits of our spectroscopy.

There are two stars, indicated by red circles, whose Gaia data suggest they are non-members,
but whose spectroscopy proves otherwise: NGC~3603-C (O3If*/WN6) and NGC~3603-38 [O3.5V((f))]. There is no question of actual membership, and both of these stars are in the crowded center.  Their RUWE values are 1.6 and 2.1, both indicative that the Gaia data are unreliable, but below our more generous cutoff of 2.5.  There are no stars whose Gaia data indicates membership but which are contradicted by our spectroscopy.

There are, of course, many stars without complete data.  There are 159 stars with neither reliable Gaia data nor spectroscopy, shown by cyan triangles.  For these stars,
we can only go by their locations in the CMD to guess membership.  There are 211 stars, shown by black triangles,  whose Gaia data suggest that they are members, but which lack spectroscopy.   A few of these are found outside the color range in which we expect to find members; either their
photometry or Gaia data are likely off.

A comparison of the figure with the data in Table~\ref{tab:photspec} shows that the identification of cluster members is increasingly muddled
at magnitudes fainter than $V=16.5$.   We find that by $V\sim17$  Gaia is identifying a few stars as cluster members that are likely too blue to belong to the cluster.  Worse, however, is the mixture of a large sample of stars without good Gaia data, and probable members and non-members at $V=17$.  Our spectroscopy becomes increasingly incomplete below $V=16$, as is readily apparent
from a comparision with Table~\ref{tab:photspec}.  The faintest stars with spectroscopy are a couple of B dwarfs at $V=16.8$; our spectroscopy is relatively complete above $V=16.0$.  
Further spectroscopy is clearly needed at $V>16$ within the expected color-band ($0.9>B-V<1.7$) but what can we answer from the copious data in hand?

One further complication we can infer from the CMD is that the extinction is not uniform within the cluster.  To a good approximation, all the spectroscopically confirmed stars will have the same intrinsic colors, to within 0.05 or less in $B-V$.  Yet the colors of the cluster
members at $V=15$ extend from $B-V\sim1.0$ to 1.7.   \citet{2008AJ....135..878M} found $(B-V)\sim 1.4$ with little variation across the cluster core (in agreement with previous studies), which would correspond roughly to $E(B-V)= 1.2$, consistent with where
the majority of the members in the CMD is found.   We re-examine this issue in the following section.

\subsection{Reddening}
\label{Sec-extinct}

Within the cluster core, \citet{2008AJ....135..878M} adopt a two-component model  
for the ratio of selective-to-total extinction, $R_V$, following the comprehensive study of \citet{2000PASJ...52..847P}, who 
adopted the standard $R_V=3.1$ for a foreground reddening of $E(B-V)$=1.1, and a value of $R_V=4.3$ for color excess above
that; i.e., $A_V=3.41+4.3[E(B-V)-1.1]$.  \citet{2008AJ....135..878M} found this consistent with their spectroscopic parallaxes (i.e.,
$M_V$ computed from the dereddened photometry) compared to that expected for their spectral types if a distance of 7.6~kpc was adopted.   \citet{2008AJ....135..878M} show that this distance is consistent with the kinematic distance to the cluster (see their Table 5 and discussion in their Section 4).\footnote{\citet{2023A&A...677A.175W} derive a distance of 6250$\pm$150~kpc using Gaia parallaxes for 10 stars located near the core.  This shorter distance 
was derived by excluding another 12 stars whose early O-types clearly indicate membership (e.g., their Table B-1) and whose proper motions indicate near-certain membership by our own analysis here.  Doubtless the  \citet{2008AJ....135..878M} distance determination will not be the last word on this subject, but we adopt their value since it agrees with the kinematic distance to the cluster.  We note that the difference would amount to $\sim$0.5~mag in the distance modulus, which
we believe would be hard to reconcile with the expected absolute magnitudes.  We refer the reader to the recent analysis of the double-lined binary NGC~3603-A1 which found an absolute magnitude of the system $M_V=-7.3$, in agreement with the $M_V=-7.3$ value derived using the 7.6~kpc distance \citep{MasseyA1}.  To reconcile this
with the 6.3~kpc distance would require stellar radii inconsistent with the light curve.}

There have been several pertinent studies over the following years based upon analysis of the gas. \citet{2016AJ....151...23P} used
archival HST/WFC3 images to determine a pixel-to-pixel distribution of the color excess $E(B-V)$ of the gas based on the H$\alpha$ to Pa$\beta$ flux ratio within the central 2\farcm4 $\times$ 2\farcm1 region of the cluster, finding $E(B-V)$ ranging from 1.5 to 2.2 mag, although, as they note, this value represents an upper limit to the reddening of the associated stars since only some of the gas will be in front of the stars.  A more recent study with JWST by \citet{2024A&A...688A.111R} also measured the color excess and $R_V$ value using recombination lines of the gas, finding $R_V=4.8\pm1.1$ but a color excess corresponding to $E(B-V)=0.64\pm0.3$.  Such a low value is inconsistent with what we know of the reddening from stars (e.g., \citealt{2000PASJ...52..847P, 
2008AJ....135..878M} and references therein). 

Using our extensive spectroscopy we can re-examine the issues of the extinction and the distance here.  We begin by determining the color excesses $E(B-V)$ as a function of position within the cluster.  The use of O-type stars for this is ideal, as their temperatures are so high that there is little variation in intrinsic color with spectral subtype.  For this
exercise we adopt the $(B-V)_0$ colors of \citet{2006A&A...457..637M} based upon the spectral energy distributions (SEDs) computed with {\sc cmfgen} models by \citet{2005A&A...436.1049M}.  We note that their analysis results in $(B-V)_0$ about 0.04~mag redder than the commonly used values (c.f.,  \citealt{Fitz,Allen}), but are likely more
accurate as older values are based on higher surface gravity models than realistic (see discussion in \citealt{2006A&A...457..637M}). According to these studies, all O stars have
an intrinsic color of $-0.27\pm0.01$, with the exact value dependent upon spectral subtype and luminosity class.  We ignored the handful of stars
with spectral types later than B1, and adopted $(B-V)_0=-0.19$ for the B1~Iab star Sh~25 \citep{Fitz} and $-0.26$ for the B0-1~V classes.  We also ignored the O3If*/WN6 stars,
as their colors will be affected by their strong emission lines.  

\begin{figure}
\plotone{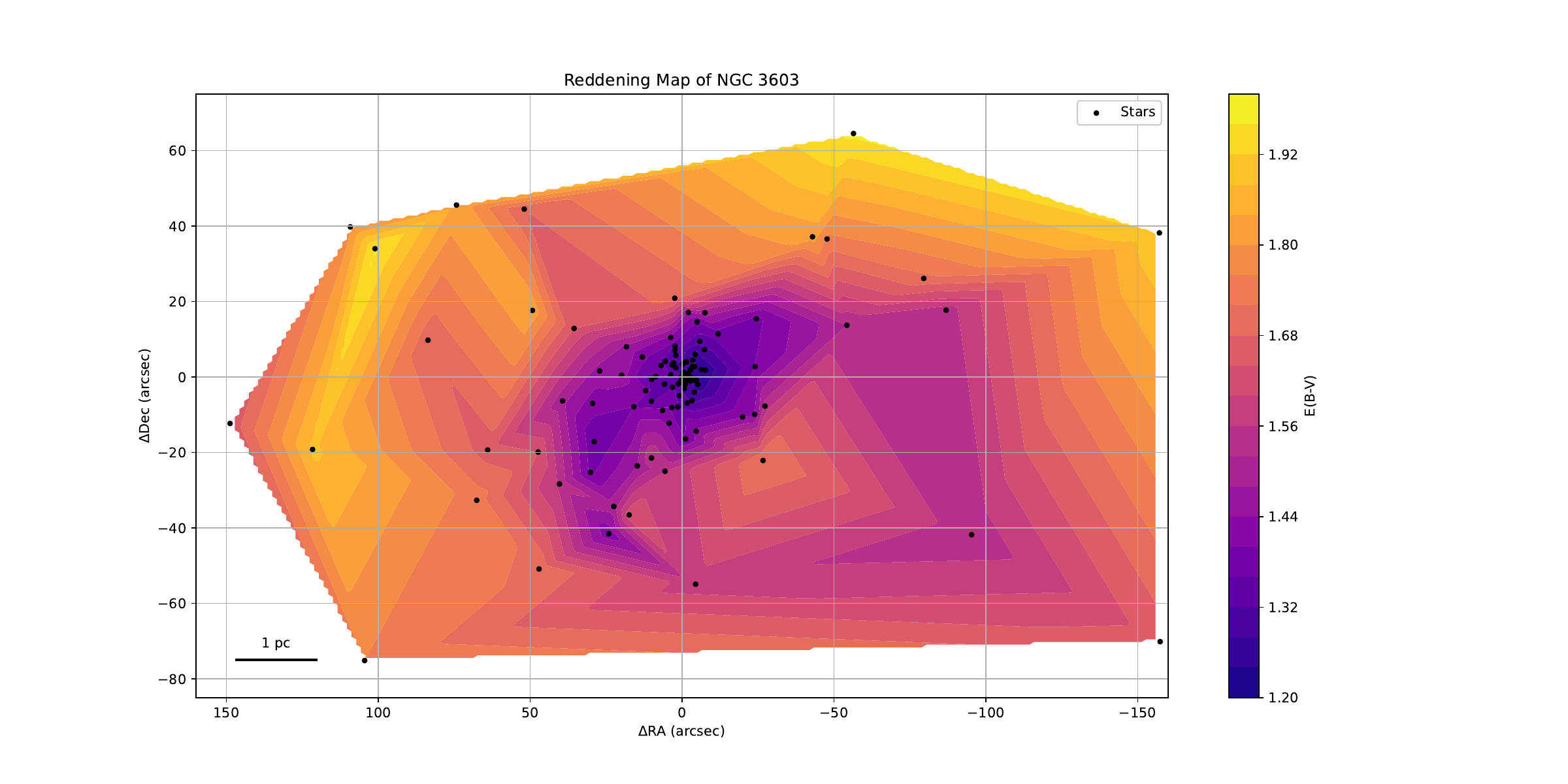}
\plotone{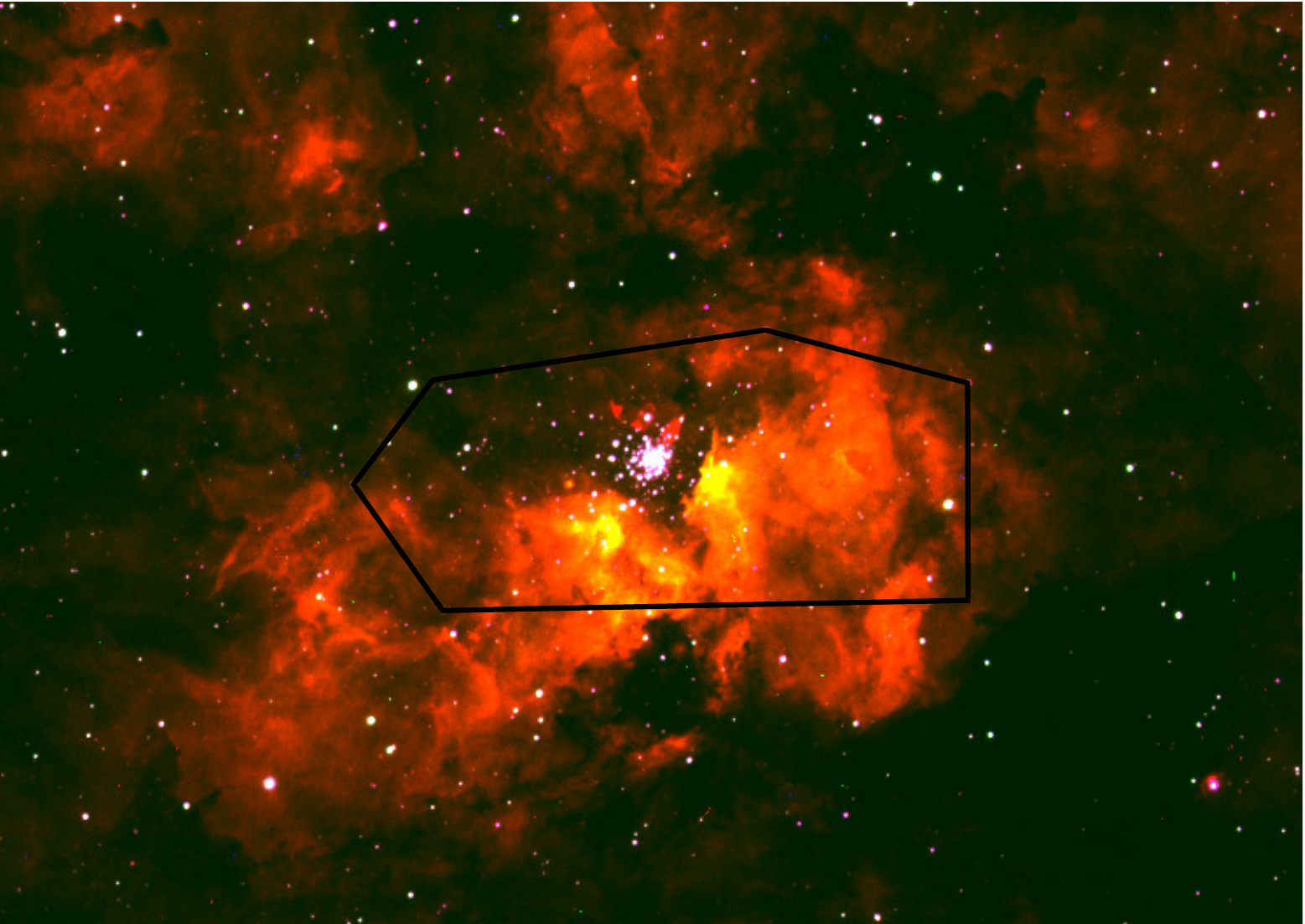}
\caption{\label{fig:ebmv} Reddening map of NGC~3603.  {\it Upper:}  This figure shows a smoothed contour map of the $E(B-V)$ values determined from the photometry of the
stars spectroscopically classified as O-type or early B.  The axes show the distance in arcseconds from the center of the cluster, taken to be NGC~3603-A1.
Note that the cluster center is located in a reddening ``hole," presumably blown by the strong stellar winds of the numerous
hot luminous stars present.  Reddening increases towards the periphery of the region.  {\it Lower:} For comparison, we show a color-rendition of the region.  The image is the 
composite of a 300 s H$\alpha$ on-band exposure (red), a 150 s H$\alpha$ continuum exposure (blue), and a 300 s O{\sc iii} $\lambda 5007$-on band exposure (green).  The
data were taken by graduate student Kennedy Farrell and co-author N. M. on UT 2025 Jan 6 on the Las Campanas 1 m Swope telescope at the end of the night during time assigned for an unrelated project.}
\end{figure}

We show a smoothed reddening map  in Figure~\ref{fig:ebmv}.  We see at once that the cluster center has by far the lowest reddening values, with values generally
increasing from $E(B-V)\sim1.2$ (deep purple) to $E(B-V)\sim2.0$ (light yellow).  The obvious interpretation is that the strong stellar winds of the myriad of hot, luminous
stars have blown a hole in the gas.  This impression is confirmed by the visual appearance of the cluster in H$\alpha$ and O\,{\sc iii} shown in the lower panel of the figure. We will discuss the age of the
cluster in more detail in the following section, but with the presence of so many short-lived O3I*/WN6 and O3 stars, the age must be close to 1-3~Myr.  So, this clearing has taken place in a remarkably short time.

\section{The H-R Diagram (HRD)}
\label{Sec-HRD}
We have already noted that there is only a negligible difference in the intrinsic $(B-V)_0$ colors of O-type stars, regardless whether they are early-type O-stars with effective temperatures
of 45,000-50,000, or late-type O-stars with effective temperatures of ``only" 30,000 K.  This degeneracy of intrinsic colors with effective temperature
is due to the fact that the flux in such stars peaks in the far-UV, and there is little change in the slope of the tail of the Rayleigh-Jeans distribution.  This degeneracy extends even
into the near-UV colors; see examples and discussion in \citet{MasseyGilmore}. Yet over this same range in temperature the visual bolometric correction changes by 1.5~mag, or a factor of 4 in luminosity.  The mass-luminosity
relationship is fairly flat for such high mass stars: we calculate that $L\sim M^{1.9}$ to $M^{2.2}$ near the zero-age main-sequence of the Z=0.014 Geneva evolutionary tracks \citep{Sylvia}, with the
smaller exponent appropriate to the highest masses.  Thus an error of a factor of 4 in luminosity translates into an error of a factor of 2 in mass, which is significant if
our interest is in measuring mass functions.

It is for this reason that we have relentlessly pursued our spectroscopic survey of the NGC~3603 stars.  The optical spectrum is quite sensitive to the effective temperature, and
the bolometric correction is easily computed from that 
temperature\footnote{From their {\sc fastwind} analysis of LMC OB stars, \citet{MasseyTeff} 
proposed that a good approximation for the bolometric correction (BC) is ${\rm BC}=-6.90\log{T_{\rm eff}} + 27.99$.  
Using {\sc cmfgen} models computed with solar metallicities,  \citet{2005A&A...436.1049M} proposed instead 
${\rm BC} =-6.80 \log{T_{\rm eff}}+27.58$. The two approximations agree to within 0.06 mag over the range $T_{\rm eff}$=20,000-50,000.
Since {\sc cmfgen} includes a more exact treatment of blanketing, and is based on a more appropriate metallicity for NGC~3603, we adopt the latter relationship here.}.  
Combined with the absolute visual magnitude $M_V$ (which, of course, depends on a knowledge of the distance and a correction for interstellar reddening) this allows us to compute an accurate luminosity, which can be used with evolutionary tracks to approximate the star's mass.

Of course, we do not have spectral types for all the stars in our sample, despite the many years of observations reported here.  In this section we will lay out
how we obtained effective temperatures and bolometric luminosities both for the stars with and without spectroscopy.

\subsection{Stars with Spectroscopy}
\label{Sec-HRDSpect}

For an accurate placement in the H-R diagram we need both the effective temperatures and the bolometric luminosities.  For the stars with
spectra, we can either obtain effective temperatures ($T_{\rm eff}$) by modeling the spectra, or infer $T_{\rm eff}$
simply from the spectral classifications.
To obtain the luminosities we compute the absolute visual magnitude $$M_V=V-A_V-14.40,$$ where 14.40 is the true distance modulus corresponding to a distance of 7.6~kpc.  The correction for interstellar extinction $A_V$ is determined from the color excess $E(B-V)$ using
the two-component model
$$A_V=3.41+4.3[E(B-V)-1.1],$$ as discussed above. 
The bolometric correction is  just $${\rm BC}=-6.80 \log{T_{\rm eff}}+27.58,$$  based on the study of \citet{2005A&A...436.1049M} as discussed above.  The bolometric
magnitude is $$m_{\rm bol} = M_V + {\rm BC},$$  and the $\log$ of the bolometric luminosity relative to the sun is computed as
$$\log L/L_\odot = (m_{\rm bol}-4.74)/-2.5,$$ where 4.74 is the bolometric magnitude of the Sun.

\subsubsection{Physical Parameters from Modeling}
\label{Sec-modeling}

For NGC~3603 there are complications in addition to the usual complexities of modeling spectra.  The main one is that because of the high
extinction, no UV data are available to establish the terminal velocities of the wind.  Lacking this information, we computed our models
adopting a value for the terminal velocity that is 2.6$\times$ the escape speed $v_{\rm esc}$ \citep{2000ARA&A..38..613K}, where the
calculation of $v_{\rm esc}$ is based on the model's surface gravity $g$, radius (computed to match the absolute visual magnitude
and effective temperature), and Eddington factor.

We modeled the optical spectra of the O stars using the  version 10 of {\sc fastwind} \citep{1997A&A...323..488S}.  This version 
includes approximate line blankening \citep{2002A&A...396..949H,2005A&A...435..669P}.
Although the {\sc fastwind} code was eventually
updated to provide a more exact treatment of the background elements  \citep{2017IAUS..329..435P}, this older version had been found to give adequate agreement
with the ``gold-standard" {\sc cmfgen} code  \citep{1998ApJ...496..407H,2003IAUS..212...70H,2012IAUS..282..229H} as shown by \citet{2013ApJ...768....6M}.

The modeling was undertaken independently by two REU students, coauthors M.H. and C.O., with some stars in common to test consistency.
The basic process is laid out in \citet{2004ApJ...608.1001M}.
Adequate fits were all found with an adopted He/H number ratio of 0.1.  In most cases, adopting the standard value $\beta$=0.8 for the wind
acceleration parameter was adequate; the one exception is noted in the table.  The models were run without clumping, and thus the tabulated mass-loss rates should be corrected downwards by a factor of $\sqrt{f}$, where $f$ is the filling factor, usually taken to be 0.1. (In other words, the values should
be decreased by roughly a factor of 3.)   We note that
the goal of this procedure was aimed at obtaining better effective temperatures than could be derived from just the spectral types, and not as a definitive analysis of their mass-loss rates or radii. This was particularly important for the earliest spectral types, as the classification scheme becomes degenerate with effective temperatures as the He\,{\sc i} lines weaken into the noise. (See discussion in \citealt{2004ApJ...608.1001M} and \citealt{MasseyTeff}.)  

The results of this modeling effort are given in Table~\ref{tab:modeling}.  We include the results from both modelers in order to show the excellent
agreement despite the subjective procedure of judging the best fits by eye.  The one exception is for the O2-3III star NGC~3603-205, where the lead
author judged the results of the two fits.

\startlongtable
\begin{deluxetable}{l l c c c c r r c c c r}
\tabletypesize{\scriptsize}
\tablecaption{\label{tab:modeling} Modeling of NGC~3603 Stars}
\tablehead{
&
&
&\multicolumn{4}{c}{Modeler 1 (CO)}
&
&\multicolumn{4}{c}{Modeler 2 (MH)} \\ \cline{4-7} \cline{9-12}
\colhead{Star}&\colhead{Sp.Type}&\colhead{Teff(Adopted)}&
\colhead{Teff}
&\colhead{$\log g$}
&\colhead{$\dot{M}$\tablenotemark{a}}
&\colhead{$v\sin{i}$\tablenotemark{b}}
&
&\colhead{Teff}
&\colhead{$\log g$}
&\colhead{$\dot{M}$\tablenotemark{a}}
&\colhead{$v\sin{i}$\tablenotemark{b}} 
}
\startdata
16   & O3V((f))      &   47750 &   47500 &     4.0 &     3.0 &     180 &&   48000 &     3.9 &     1.5 &     180 \\
38   & O3.5V((f))    &   43000 &   43000 &     3.8 &     1.6 &     150 &&   43000 &     3.7 &     0.9 &     160 \\
42   & O3.5III       &   44000 & \nodata & \nodata & \nodata & \nodata &&   44000 &     3.8 &     2.5 &     120 \\
101  & O6.5Vz        &   39000 &   39000 &     4.0 &     0.2 &     180 && \nodata & \nodata & \nodata & \nodata \\
102  & O8.5V/O9c     &   35000 &   35000 &     4.0 &     0.2 &     150 && \nodata & \nodata & \nodata & \nodata \\
103  & O3III((f*))   &   45750 &   45000 &     3.8 &     2.5 &     135 &&   46500 &     3.8 &     1.2 &     170 \\
108  & O4.5V         &   42000 &   42000 &     4.0 &     0.8 &     110 && \nodata & \nodata & \nodata & \nodata \\
109  & O8V           &   37500 &   37500 &     4.0 &     0.2 &     180 && \nodata & \nodata & \nodata & \nodata \\
104  & O3III((f*))   &   48500 & \nodata & \nodata & \nodata & \nodata &&   48500 &     3.9 &     2.5 &     150 \\
116  & O3.5V((f))    &   45000 &   45000 &     4.0 &     0.2 &     150 && \nodata & \nodata & \nodata & \nodata \\
117  & O5.5V((f))z   &   40000 &   40000 &     4.0 &     0.4 &     330 && \nodata & \nodata & \nodata & \nodata \\
120  & O6V           &   40750 &   40500 &     4.0 &     0.2 &     110 &&   41000 &     4.0 &     0.2 &     130 \\
124  & O6V           &   37500 &   37500 &     3.8 &     0.2 &     130 && \nodata & \nodata & \nodata & \nodata \\
125  & O6V((f))      &   39500 &   40000 &     4.0 &     0.3 &     150 &&   39000 &     3.9 &     0.0 &     150 \\
128  & O9V           &   34000 &   34000 &     4.1 &     0.8 &     100 && \nodata & \nodata & \nodata & \nodata \\
135  & O8V           &   36000 & \nodata & \nodata & \nodata & \nodata &&   36000 &     3.9 &     0.2 &     280 \\
141  & O8.5V         &   44000 &   44000 &     4.1 &     4.0 &     150 && \nodata & \nodata & \nodata & \nodata \\
205  & O2-O3III      &   43500 &   43500 &     3.9 &     1.6 &     120 &&   52000 &     4.0 &     2.0 &     120 \\
210  & O7V           &   37000 & \nodata & \nodata & \nodata & \nodata &&   37000 &     3.8 &     0.2 &     230 \\
214  & O5.5V         &   39000 & \nodata & \nodata & \nodata & \nodata &&   39000 &     3.7 &     0.2 &     115 \\
215  & O7V           &   37000 &   37000 &     4.0 &     0.7 &     200 && \nodata & \nodata & \nodata & \nodata \\
216  & O7.5V         &   36000 &   36000 &     4.0 &     0.1 &     260 && \nodata & \nodata & \nodata & \nodata \\
218  & O6.5V         &   37000 &   35500 &     3.4 &     0.2 &     130 &&   38500 &     4.0 &     0.2 &     120 \\
219  & O8V           &   37000 &   37000 &     4.0 &     0.2 &     300 && \nodata & \nodata & \nodata & \nodata \\
220  & O7.5V         &   36000 &   36000 &     4.0 &     0.5 &     400 && \nodata & \nodata & \nodata & \nodata \\
222  & O9.7V         &   34000 & \nodata & \nodata & \nodata & \nodata &&   34000 &     4.0 &     0.0 &     120 \\
227  & B0.5V         &   31000 &   31000 &     4.1 &     0.5 &     250 && \nodata & \nodata & \nodata & \nodata \\
305  & O3.5V         &   44250 &   42000 &     3.8 &     0.5 &     200 &&   46500 &     4.0 &     0.4 &     170 \\
308  & O6V((f))      &   40500 & \nodata & \nodata & \nodata & \nodata &&   40500 &     4.0 &     0.2 &     300 \\
310  & O5V           &   45000 &   45000 &     4.0 &     2.0 &     125 && \nodata & \nodata & \nodata & \nodata \\
313  & O6V           &   38500 & \nodata & \nodata & \nodata & \nodata &&   38500 &     3.9 &     0.1 &     300 \\
316  & O3.5V         &   46000 &   46000 &     3.9 &     2.0 &     125 && \nodata & \nodata & \nodata & \nodata \\
319  & O3.5V         &   46000 &   46000 &     4.0 &     1.5 &     170 && \nodata & \nodata & \nodata & \nodata \\
320  & B0-B0.2V      &   30000 &   30000 &     4.1 &     0.1 &     160 && \nodata & \nodata & \nodata & \nodata \\
334  & O9.5V         &   33500 &   33500 &     3.8 &     0.2 &     160 && \nodata & \nodata & \nodata & \nodata \\
Sh17 & O5.5III       &   39750 &   40000 &     3.8 &     0.8 &     135 &&   39500 &     3.9 &     1.7 &     140 \\
Sh19 & O3V((f))      &   45500 & \nodata & \nodata & \nodata & \nodata &&   45500 &     3.8 &     1.0 &     175 \\
Sh20 & O9V           &   36000 &   36000 &     4.0 &     0.2 &     100 &&   36000 &     4.0 &     0.1 &      90 \\
Sh22 & O3III(f)      &   48500 & \nodata & \nodata & \nodata & \nodata &&   48500 &     4.0 &     5.0 &     200 \\
Sh23 & OC9.7Ia       &   29000 & \nodata & \nodata & \nodata & \nodata &&   29000 &     3.1 &     2.2\tablenotemark{c} &     150 \\
Sh47 & O3.5III(f)    &   47000 & \nodata & \nodata & \nodata & \nodata &&   47000 &     4.1 &     2.5 &     210 \\
Sh48 & O5.5V         &   40000 & \nodata & \nodata & \nodata & \nodata &&   40000 &     3.7 &     0.7 &     240 \\
Sh50 & O8.5III       &   36000 & \nodata & \nodata & \nodata & \nodata &&   36000 &     3.9 &     0.2 &     170 \\
Sh52 & O7V           &   36500 & \nodata & \nodata & \nodata & \nodata &&   36500 &     4.0 &     0.0 &     200 \\
Sh57 & O3III(f)      &   47000 & \nodata & \nodata & \nodata & \nodata &&   47000 &     3.9 &     4.0 &     145 \\
Sh59 & O8III         &   37000 &   37000 &     4.1 &     0.8 &     100 &&   37000 &     4.0 &     0.6 &      95 \\
Sh63 & O3.5III((f))  &   44500 & \nodata & \nodata & \nodata & \nodata &&   44500 &     3.7 &     0.8 &     170 \\
\enddata
\tablenotetext{a}{Units of $10^{-6}M_\odot$ yr$^{-1}$ computed with a filling factor $f$ of 1.0. Actual mass-loss rates
should be adjusted down by a factor of $\sim$3 for standard clumping $f=0.1$.}
\tablenotetext{b}{Projected rotational velocity in units of km s$^{-1}$.}
\tablenotetext{c}{Model computed with $\beta=1.2$.}
\end{deluxetable}

\subsubsection{Physical Parameters from Spectral Types}

Not all of our spectra lent themselves to modeling; some spectra lacked the high signal-to-noise that accurate modeling requires, and in other cases the wavelength coverage did not include H$\alpha$, crucial
for determining the mass-loss rate $\dot{M}.$  In particular the O2-3I*f/WN5-6 ``slash" stars required more sophisticated treatment than easily
achieved with {\sc fastwind}.  For those stars, we adopted a value of 42,000~K, following the modeling of the (composite) spectrum of the NGC~3603-A1 binary (\citealt{2010MNRAS.408..731C}; see also \citealt{MasseyA1}).  For the other O-type stars that were not modeled, but had spectral types, we relied upon the theoretical calibration of spectral
type and effective temperature by \citet{2005A&A...436.1049M} using {\sc cmfgen} models.  Their results
are  in good agreement with the effective temperature scale found by \citet{MasseyTeff}, derived by fitting LMC O-type stars with version 10 of {\sc fastwind}.   

For the B dwarfs, the temperature scale is much less well established. The most modern effort is that of \citet{2007A&A...471..625T}, who applied a grid of TLUSTY to newly classified B stars as part of a large spectroscopic survey of massive stars. The values in their Table 10 only
include early dwarfs for the Milky Way, and their B0~V temperature is actually hotter than what we adopt for an O9.5~V star.  To avoid this problem,
we arrived at a compromise scale between their work and the scale given in Table 2 of  \citet{2006ApJ...648..580H}.  Those values trace to \citet{1997ApJ...489..698H}, which in turn got them from  \citet{1989AJ.....98.1305M}, which was mostly based on \citet{1977A&A....54...31F}. We propose that the problem with B dwarfs is two fold. First, unlike the O-type
stars, the spectral classification criteria are somewhat poorly defined in terms of the relative strengths of Si\,{\sc ii} vs. Si\,{\sc iii} vs. Si\,{\sc iv} line strengths.  The problem is exacerbated by these lines being weaker at higher surface gravities (i.e., lower luminosities). Secondly, the modeling of
these metal lines is certainly more complex than the case for the O stars, where the classification criteria are based {bf primarily} on the relative
strengths of He\,{\sc i} and He\,{\sc ii}: helium is a much simpler atom. 
The situation for our one B-type supergiant is considerably better, due to the {\sc cmfgen} modeling work of \citet{2006A&A...446..279C}.
For our B1~Iab star Sh 25, we adopt an effective temperature of 21,000~K based on their work. (See also \citealt{2008ASPC..388..109C} and \citealt{2023A&A...677A.175W}.  The latter paper argues that Sh 25 is coincidentally a foreground star; we discuss this possibility further
below in Section~\ref{Sec-tracks}.) 

We remind the reader that while an uncertainty in the temperature by itself is mostly
parallel to the evolutionary tracks, and thus does not affect our estimation of the mass, the luminosity is affected because of the dependence
of the bolometric correction on the temperature adding to its importance.


\subsubsection{The Final Parameters for Stars with Spectral Types}

In Table~\ref{tab:HRDSpect} we list the physical properties we adopt for the member stars with spectroscopy.   We indicate whether the
parameters come from modeling or from the spectral types.  For the binaries we adopted a temperature intermediate between the two
components and computed the luminosity of the combined system, and note these cases.  (This decision is defended in Section~\ref{Sec-IMF}.)

\begin{deluxetable}{l l c c c c c}
\tablecaption{\label{tab:HRDSpect} Properties of NGC~3603 Stars with Spectra}
\tablehead{
\colhead{Star}
&\colhead{Sp.Type}
&\colhead{$E(B-V)$}
&\colhead{$M_V$}
&\colhead{$\log{T_{\rm eff}}$}
&\colhead{$\log{L/L_\odot}$}
&\colhead{Type\tablenotemark{a}}
}
\startdata
A1   & O3If*/WN6     & 1.25 & -7.28 & 4.623 & 6.35 & 3 \\
B    & O3If*/WN6     & 1.25 & -7.13 & 4.623 & 6.29 & 2 \\
C    & O3If*/WN6     & 1.25 & -6.56 & 4.623 & 6.06 & 2 \\
Sh25 & B1Iab         & 1.69 & -8.02 & 4.301 & 5.77 & 2 \\
A2   & O3V((f))      & 1.31 & -6.19 & 4.649 & 5.99 & 2 \\
Sh18 & O3.5If        & 1.56 & -7.11 & 4.619 & 6.27 & 2 \\
Sh47 & O3.5III(f)    & 1.48 & -6.72 & 4.672 & 6.26 & 1 \\
Sh23 & OC9.7Ia       & 1.46 & -6.62 & 4.462 & 5.65 & 1 \\
301  & O4Ifc         & 1.31 & -5.95 & 4.610 & 5.78 & 2 \\
42   & O3.5III       & 1.30 & -5.66 & 4.643 & 5.76 & 1 \\
104  & O3III((f*))   & 1.33 & -5.77 & 4.686 & 5.92 & 1 \\
302  & O3.5V         & 1.26 & -5.46 & 4.644 & 5.68 & 2 \\
103  & O3III((f*))   & 1.25 & -5.35 & 4.660 & 5.68 & 1 \\
A3   & O3.5III(f)    & 1.32 & -5.63 & 4.625 & 5.70 & 2 \\
303  & O3III         & 1.28 & -5.37 & 4.633 & 5.61 & 2 \\
\enddata
\tablenotetext{a}{1--$T{\rm eff}$ determined by modeling (e.g., Table~\ref{tab:modeling}); 2--$T{\rm eff}$
determined from spectral type; 3--$T_{\rm eff}$ determined by spectral type but system is binary.}
\tablecomments{Table~\ref{tab:HRDSpect} is published in its entirety in machine-readable format.
A portion is shown here for guidance regarding its form and content.}
\end{deluxetable}

\subsection{Stars without Spectroscopy}

\begin{figure}
\epsscale{0.57}
\plotone{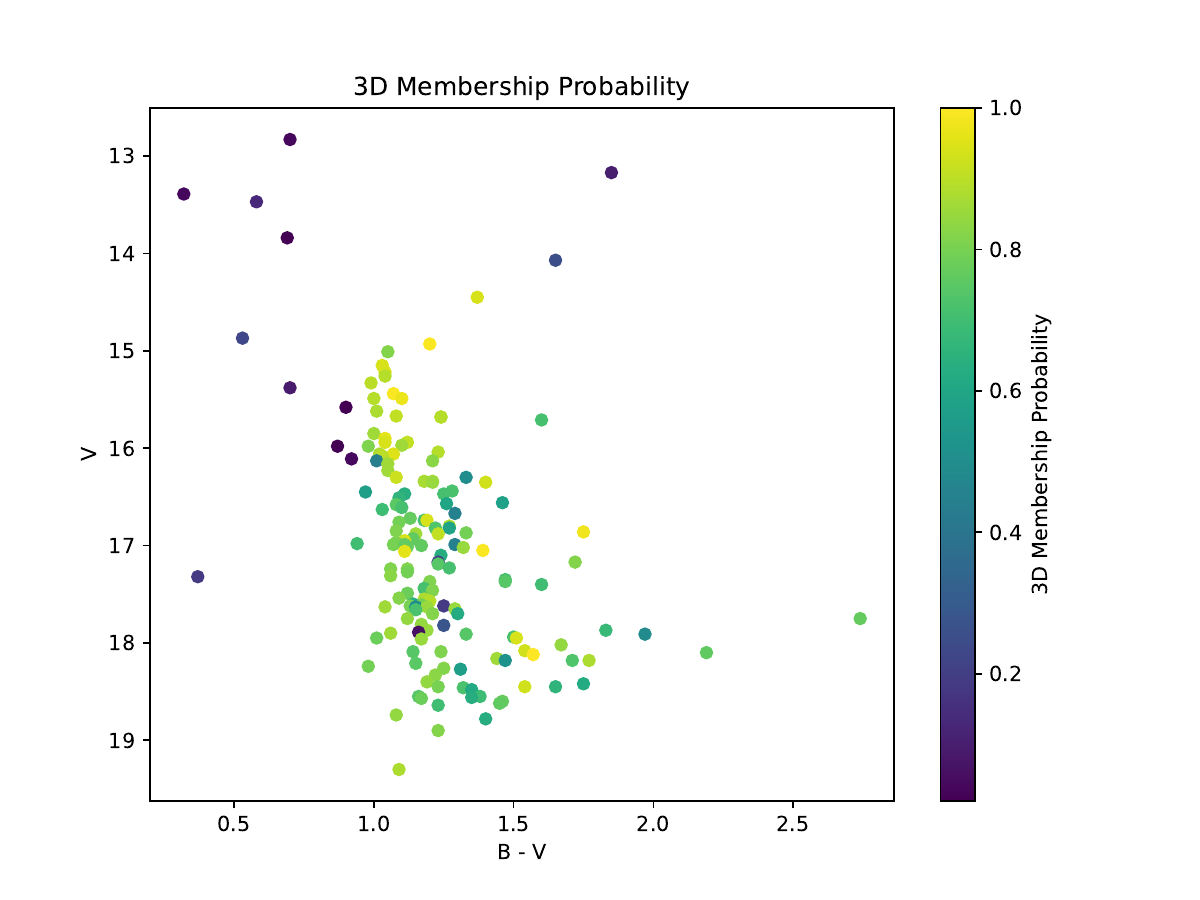}
\plotone{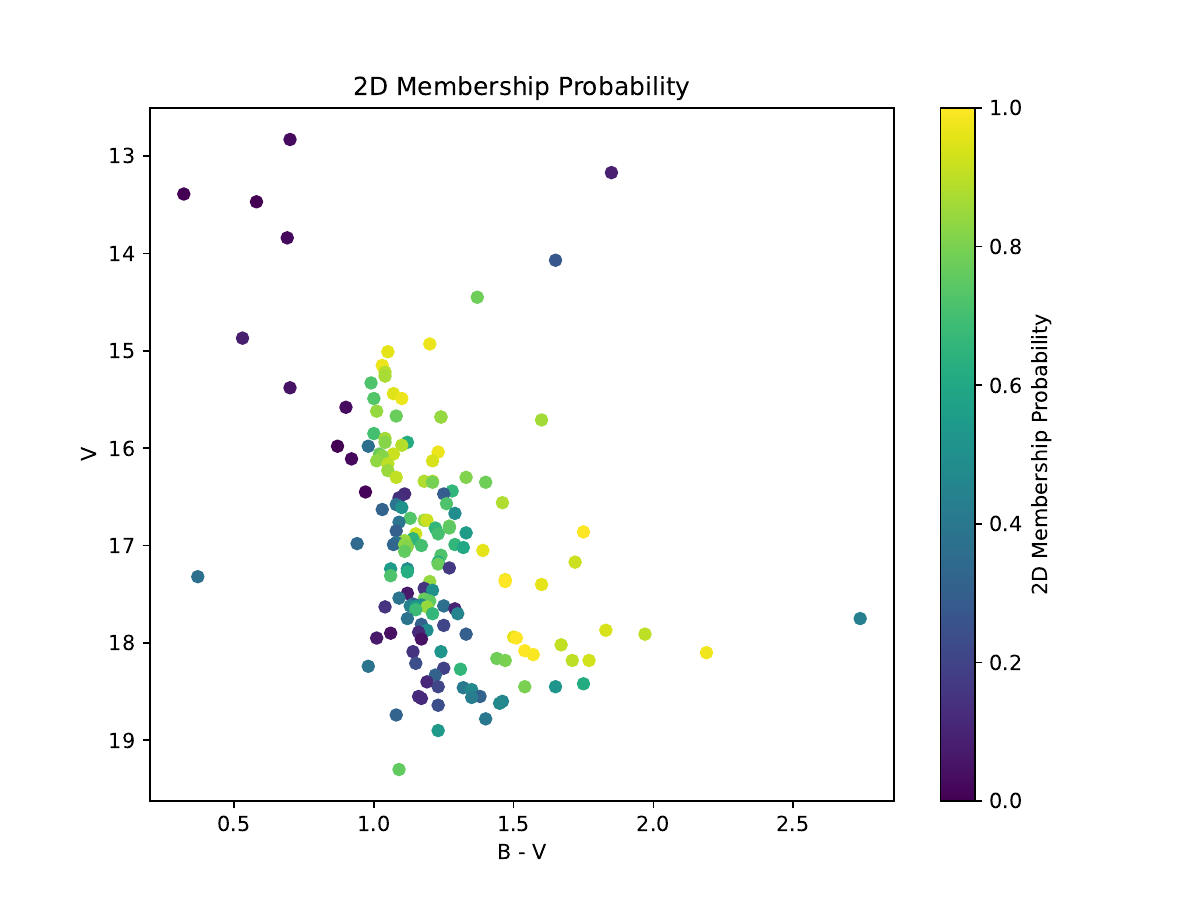}
\plotone{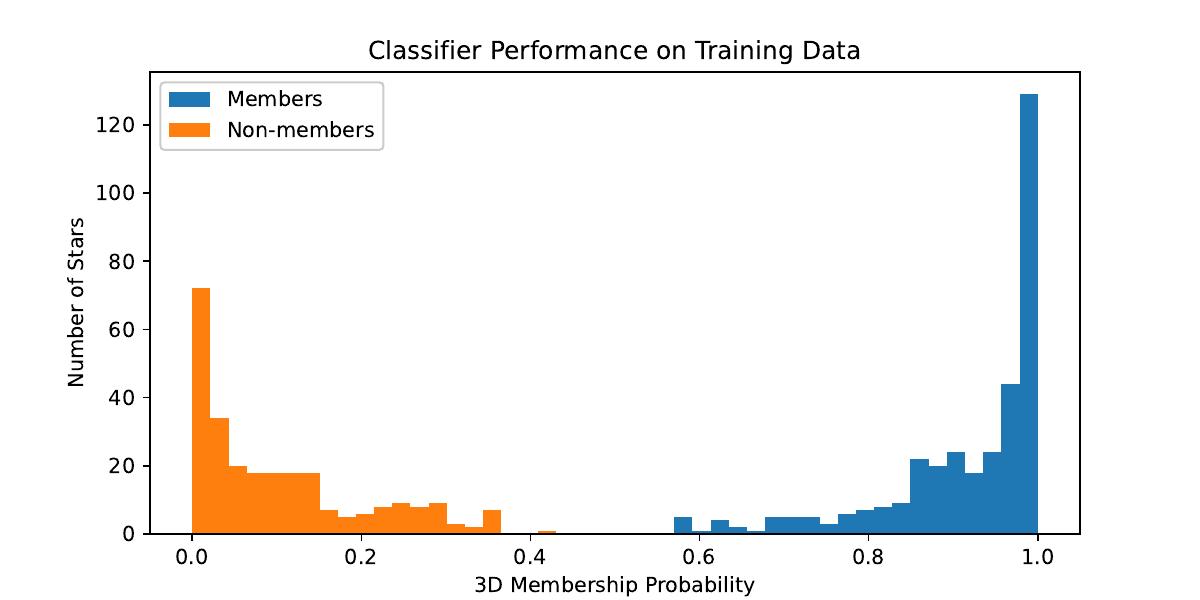}
\plotone{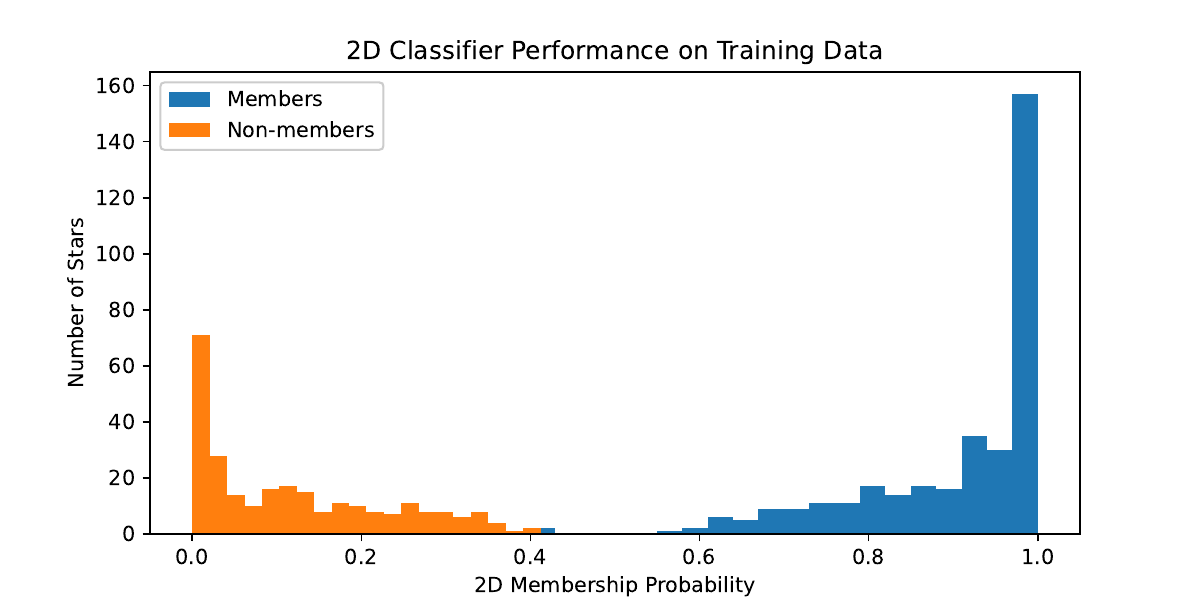}
\caption{\label{fig:forest} Membership probability for stars without spectroscopy.  {\it Upper panels:} For the stars without spectroscopy and without adequate Gaia data, we use a random forest machine-learning classifier to assign probabilities of membership based upon their location in the CMD.  On the {\it left} we also include the star's radial distance from the center of the cluster in
assigning probabilities. On the {\it right} we use only the magnitude and color information.  The training data consists of the information on known members and known nonmembers.  {\it Lower panels:} Results of applying the classifier on the training set itself as a reality check.}
\end{figure}

There are two problems for the stars without spectral types.  The first of these is determining a membership probability for the stars without Gaia data.  For the stars with spectroscopy, the Gaia data were somewhat superfluous as a chance alignment of a rare O-type star with
the cluster is negligible. However, for the stars without spectroscopy or Gaia data we must determine the probability of membership.
There are 214 stars without spectroscopy but whose Gaia data suggests they are likely members, usually with probabilities of 1.0.
However, there are 143 stars without spectroscopy that either lack Gaia data or whose
Gaia data are too poor to use (i.e., RUWE $>$2.5).   In order to address those, we utilized a ``random forest" machine learning classifier on the stars' photometry along with their radial distance from the cluster's center. We used the data of the known members and known non-members for
training. The corresponding 3D
probabilities as a function of location in the CMD are shown in the upper left panel of Figure~\ref{fig:forest}.   As a reality check, we also ran the training data through the algorithm to see how  clean the resulting separations are.  The results are shown in the lower panel of the figure.

This approach has the problem that there are few known members and nonmembers at faint magnitudes to provide guidance to the algorithm, and the classifier therefore relies heavily on the radial distance.  However, there are few faint members or nonmembers known near the center of the cluster owing to the effects of crowding on the Gaia data,
and our need to focus our spectroscopy on the brighter members.  Thus there is a built-in bias towards assigning membership to stars with small radial distances.   Applying the classifier using only the photometry
(and not the radial distance) produces a much more pessimistic assessment of membership.  We show the results of this 2D analysis on the right side of Figure~\ref{fig:forest}. We retain both sets of probabilities in our analysis.

The second problem is how to assign physical properties to these stars. If we assume that the cluster is coeval, 
then the absolute visual magnitude $M_V$ can be used to estimate the effective temperature. We use the reddening map in Figure~\ref{fig:ebmv} to approximate the $E(B-V)$ for all the stars without spectroscopy but whose Gaia data or location in the CMD 
suggests a probability of membership $>$50\%.    This allows us to compute the absolute visual magnitude $M_V$.

The vast majority of stars without spectroscopy (blue or black triangles in Figure~\ref{fig:cmd}) are fainter
than the high mass stars with the spectroscopy we've been discussing, and a reasonable assumption is that they are all dwarfs.  For dwarfs we find an accurate
relationship between effective temperature ($T_{\rm eff}$) and absolute visual magnitude ($M_V$):
$$T_{\rm eff} = 10605-3061.6 \times M_V +516.53 \times M_V^2,$$ where we have extended the relationship between $T_{\rm eff}$ and $M_V$ for O-type
stars from \citet{2005A&A...436.1049M} to late B stars using the values given in Table 15.7 of \citet{Allen}.  The scatter on that relationship is 450~K.  Similarly we use the BSTAR2006 models \citep{2007ApJS..169...83L} with the {\sc cmfgen} models \citep{2005A&A...436.1049M} to obtain an approximate relationship between the bolometric correction (BC)
and absolute visual magnitude:  $${\rm BC}=-0.06+0.60 \times M_V.$$ The scatter on that relationship is 0.04~mag.

We emphasize that these relations are approximate, and rely heavily on the assumption that these stars with only photometry are located near the zero-age
main-sequence.  We list the derived physical parameters in Tables~\ref{tab:HRDPhotgaia} and ~\ref{tab:HRDPhotNogaia}.

\begin{deluxetable}{l c c c c c c c c}
\tablecaption{\label{tab:HRDPhotgaia} Properties of NGC~3603 Stars without Spectroscopy but with Gaia Data}
\tablehead{
&
&
&
&\multicolumn{2}{c}{Reddening}
&
&
& \\ \cline{5-6}
\colhead{Star}&\colhead{$V$}&\colhead{$B-V$}&\colhead{Mem. Prob.}&\colhead{$E(B-V)$}&\colhead{S\tablenotemark{a}}
&\colhead{$M_V$}
&\colhead{$\log{T_{\rm eff}}$}
&\colhead{$\log{L/L_\odot}$}
}
\startdata
2026&14.87&   0.78&  0.999&1.80&E& -5.95&4.673& 5.73 \\
2045&15.36&   1.10&  1.000&1.34&I& -3.48&4.440& 4.15 \\
2054&15.60&   1.55&  1.000&1.80&E& -5.22&4.610& 5.26 \\
2060&15.69&   1.11&  1.000&1.37&I& -3.28&4.418& 4.02 \\
2070&15.90&   1.25&  1.000&1.53&I& -3.78&4.471& 4.34 \\
2075&15.97&   1.11&  1.000&1.59&I& -3.93&4.486& 4.44 \\
1068&16.04&   1.69&  1.000&1.80&E& -4.78&4.569& 4.98 \\
2085&16.05&   1.38&  1.000&1.69&I& -4.31&4.524& 4.68 \\
2096&16.13&   1.16&  1.000&1.46&I& -3.22&4.412& 3.98 \\
2097&16.13&   1.25&  1.000&1.55&I& -3.64&4.456& 4.25 \\
\enddata
\tablenotemark{a}{Source of reddening: I=interpolated; E=extrapolated.}
\tablecomments{Table~\ref{tab:HRDPhotgaia} is published in its entirety in the machine-readable format. A portion is shown here for guidance regarding its form and content.}
\end{deluxetable}

\begin{deluxetable}{l c c c c c c c c c c}
\tablecaption{\label{tab:HRDPhotNogaia} Properties of NGC~3603 stars without spectroscopy and Without Gaia Data}
\tablehead{
&
&
&\multicolumn{2}{c}{Mem. Prob.}
&
&\multicolumn{2}{c}{Reddening}
&
&
&  \\ \cline{4-5} \cline{7-8}
\colhead{Star}
&\colhead{$V$}
&\colhead{$B-V$}
&\colhead{3D} 
&\colhead{2D}
&
&\colhead{$E(B-V)$}
&\colhead{S\tablenotemark{a}}
&\colhead{$M_V$}
&\colhead{$\log{T_{\rm eff}}$}
&\colhead{$\log{L/L_\odot}$}
}
\startdata
2020&14.45&   1.37& 0.94& 0.78&&1.40&I& -4.65&4.557& 4.90\\
2027&14.93&   1.20& 0.99& 0.97&&1.39&I& -4.15&4.508& 4.58\\
2029&15.01&   1.05& 0.82& 0.96&&1.44&I& -4.27&4.519& 4.65\\
2034&15.15&   1.03& 0.94& 0.97&&1.33&I& -3.64&4.456& 4.25\\
2037&15.22&   1.04& 0.94& 0.89&&1.30&I& -3.45&4.437& 4.13\\
2040&15.26&   1.04& 0.89& 0.87&&1.37&I& -3.73&4.465& 4.31\\
2044&15.33&   0.99& 0.90& 0.72&&1.26&I& -3.19&4.409& 3.96\\
2048&15.44&   1.07& 0.99& 0.95&&1.31&I& -3.29&4.419& 4.02\\
2050&15.49&   1.00& 0.89& 0.73&&1.25&I& -2.96&4.384& 3.82\\
2051&15.49&   1.10& 0.97& 0.97&&1.36&I& -3.44&4.436& 4.12\\
\enddata
\tablenotemark{a}{Source of reddening: I=interpolated; E=extrapolated.}
\tablecomments{Table~\ref{tab:HRDPhotNogaia} is published in its entirety in
machine-readable format.  A portion is shown here for guidance regarding its form and content.}
\end{deluxetable}

\subsection{Comparison with Evolutionary Tracks}
\label{Sec-tracks} 

\begin{figure}
\plotone{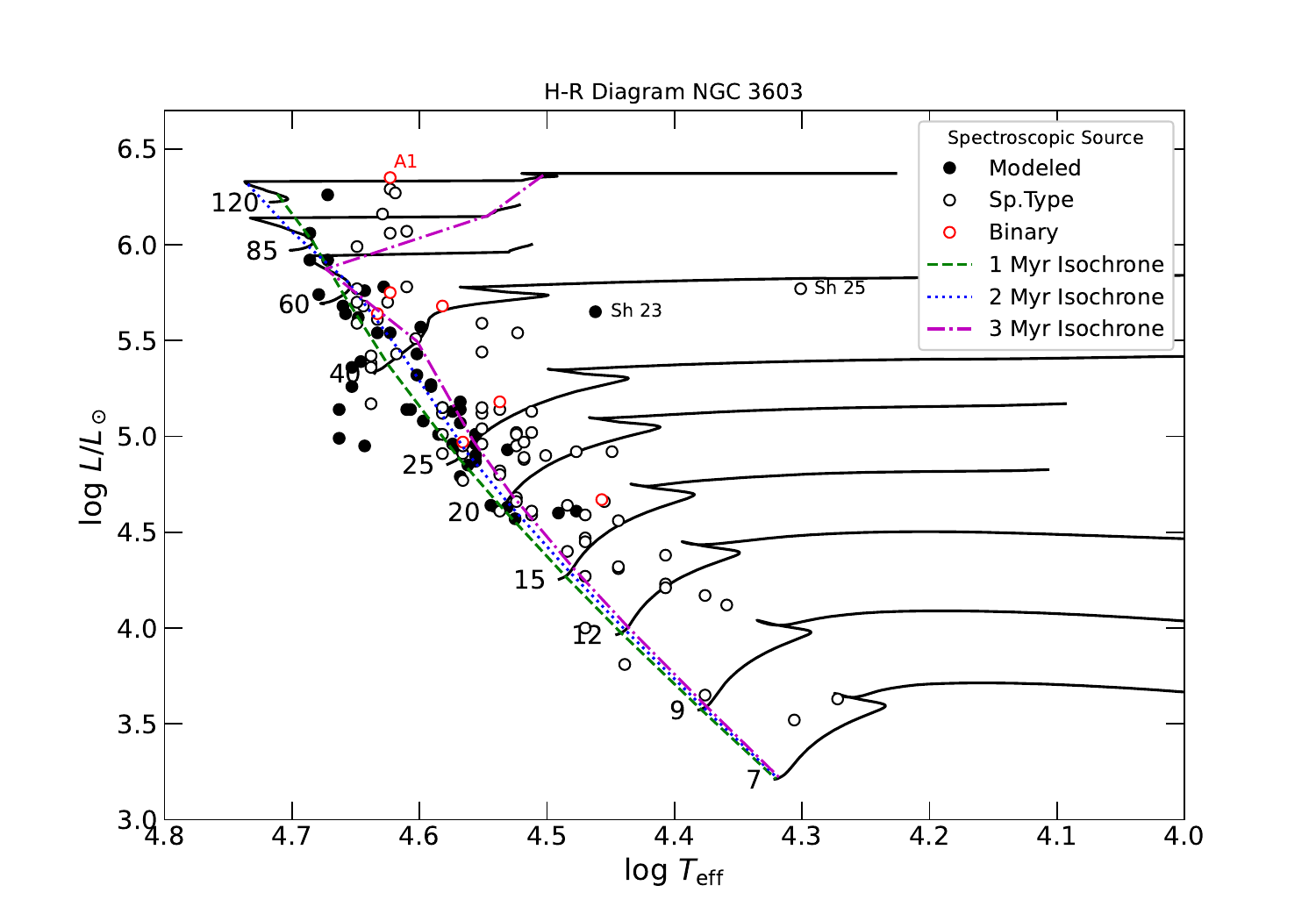}
\caption{\label{fig:HRDspec} NGC~3603 H-R Diagram for Stars with Spectroscopy.  The luminosities and effective temperatures determined with spectroscopy are plotted in this H-R diagram.  The data come from Table~\ref{tab:HRDSpect}.  The Geneva evolutionary tracks from \citet{Sylvia} are shown by solid black lines, with their initial masses labeled near the beginning of the tracks.  To simplify the figure, the tracks have been truncated.  For the three highest mass tracks ($60M_\odot$, $80M_\odot$, and $120M_\odot$) we stopped at the main-sequence turnoff (line 110 in the modes). For the others, we only show the tracks until the beginning of He-burning (line 190 in the models).  Isochrones have been computed
using linear interpolation in the models.}
\end{figure}

The goal of this project has been to count the number of stars as a function of mass, and for this we must compare
the location of stars in the HRD to evolutionary tracks in order to assign masses.  We show such a diagram in
Figure~\ref{fig:HRDspec}, where we use the Geneva evolutionary models of \citet{Sylvia} computed with rotation.  

There is the expected plethora of massive stars, consistent with our spectroscopy.  A few stars are found to the left of the zero-age main-sequence (ZAMS), presumably due to over-estimating the effects of reddening; i.e., we expect those stars have good temperatures (many were modeled)
but are simply too low in the H-R diagram.  For the most part, the ages of stars are consistent with 1-3~Myr.  

The most glaring exception is Sh 25. 
  \citet{2023A&A...677A.175W} argues on the basis of the star's luminosity that the star is in the foreground.  Their modeling
derives a $\log L/L_\odot$ of 5.48 based upon a distance derived by an intriguing new method \citep{2022A&A...668A..92W}, but one which requires an accurate assessment of the reddening (see equation 3 in \citealt{2022A&A...668A..92W}).   We do not agree with this interpretation. 
First, the chances of a B1~Iab just happening to be seen superposed on the NGC~3603 cluster is quite low.  We back this up by noting that
Sh 25 is the {\it only} B1 supergiant in our sample---they are rare.  According to the Besan\c{c}on model of the Milky Way, we expect to come across less than $10^{-2}$ OB supergiants per square degree at the Galactic latitude and longitude of NGC~3603.  The area covered by our survey is 0.09 deg$^2$, so the superposition of a foreground B supergiant is indeed statistically unlikely.  

We do agree with  \citet{2023A&A...677A.175W} that were Sh 25 a member, it would be substantially older than the majority of the other OB
stars. (They quote a value of 7~Myr but that is based on the lower luminosity they derive.)  However, this is not uncommon in otherwise coeval massive
clusters.  For instance, in their study of NGC~6611, \citet{1993AJ....106.1906H} find an age of 2$\pm1$~Myr for the vast majority of stars, and
argue that as far as the data could tell, most of the stars might have been formed on a particular Tuesday.  Despite this, there is one ``lower" mass
star (30$M_\odot$) with an age of about 6~Myr.  They make the analogy to the popping of pop corn, with a few kernels going off before the 
great burst that pops most of the corn. Regardless of the possible facetiousness of the metaphor, the presence of an occasional older star in an otherwise coeval rich, massive cluster is not unprecedented.  Besides NGC~6611, the reader is referred to the H-R diagram of Cyg OB2 shown in \citet{1991AJ....101.1408M}.  R136 itself has such an example, the O8~III star Mk~32. (See Table 1 and Figure 7 in \citealt{MH98}.  Mk~32 is the
star at $\log{T_{\rm eff}}\sim 4.56$ and $M_{\rm bol}\sim 9.6$.)   And here in NGC~3603 we find a second example: 
the OC9.7~Ia star Sh~23 is similarly a bit older.   Of course, binary evolution could be invoked to explain these anomalous members.  For the
purposes of computing the slope of the IMF it is irrelevant if we count Sh~25 as a member or not, but lacking compelling evidence to the
contrary, we include it in our numbers below.

Near the bottom of the HRD we see a curious shift of the stars towards cooler temperatures.  Are these stars misplaced?  Without exception, all stars with spectroscopy and luminosities below $\log{L/L_\odot}=4.5$  are the handful of early B-type stars we identified.  As we previously noted, the effective temperature scale of B dwarfs is not as solidly determined as that of O-type stars, but can this fully explain their peculiar location?
The two stars with the lowest assigned luminosities are 1093 (B2~III) with an adopted temperature of 20200 and 1112 (B3~V) with an assigned temperature of 18700.  To shift these stars close to the ZAMS would require their temperatures to be 20800.  This change is well within the uncertainties of both the spectral typing and the effective temperature scale once we get to the B stars.  The fact that these are systematic further suggests
that the problem is with the adopted scale.

We do not include the stars without spectroscopy in the plot, as the way we have extracted their physical properties from their photometry results in them simply paralleling the isochrones
with a handful of stars between the 25-40$M_\odot$ tracks until they become dominant at lower masses.  We will demonstrate this numerically in the next section where we consider the IMF slope.  

\clearpage
\section{The Slope of the IMF}
\label{Sec-IMF}

\citet{1955ApJ...121..161S} introduced the notion of an ``original mass function."\footnote{Although \citet{1955ApJ...121..161S} is often referenced only for the sake of the $-1.35$ value of the exponent of the mass function, this extraordinary paper did far more than that: Using the luminosity function of main-sequence stars, the study successfully tested the hypothesis that stars evolve off the main sequence after burning $\sim$10 per cent of their hydrogen mass, and demonstrated that star formation has remained relatively constant in the solar neighborhood over the past five billion years. For more, see \citet{2019NatAs...3..482K}.} Arguably, it was \citet{1979ApJS...41..513M} and \citet{1980FCPh....5..287T} who put the study of star formation rates and the initial mass function onto a modern footing. Using the notation of \citet{1980FCPh....5..287T}, the number of stars formed in the mass interval (m, m+dm) in the time interval (t, t+dt) is described as $$\phi(m) \psi(t)\ dm\ dt,$$ where $\phi(m)$ is called the IMF and $\psi(t)$ is the star formation rate (SFR), the total mass of stars formed per unit time.  $\phi(m)$ is normalized in such a way that $\int_0^{\infty} m\phi(m)\ dm = 1$. and is well described as a power-law, with $\phi(m) \propto m^{-(1+x)}$.  As \citet{1980FCPh....5..287T} notes, x, -x, (1+x), and -(1+x) have all been called the slope of the IMF. We retain the terminology used in the introduction where $\Gamma = -x$ is the slope. 
Following \citet{1986FCPh...11....1S},  it is convenient to we define the quantity $\xi(\log{m})$ as the number of stars per mass bin normalized by
the size of the bin expressed as the base-10 logarithm of the ratio of the upper and lower masses of the bin, as well as by the area expressed
in units of kpc$^2$.  Then the slope of the mass function is simply $\Gamma=d \log{\xi(\log{m})}/d \log{m}.$ As a reminder, the \citet{1955ApJ...121..161S} IMF slope has a nominal value of $\Gamma=-1.35.$  

Of course, what one actually measures is the present day mass function (PDMF). The connection between the PDMF and IMF depends on the star-forming history.  When talking about an extended region containing many star-forming regions one might
assume a continuous SFR, and adjust the counts by the relative main-sequence lifetimes as a function of mass; at the other extreme, a single cluster or OB association may be considered to be coeval (or at least ``mostly" coeval) and the slope of the IMF and PDMF the same, but note that the upper-most part of the PDMF may have been depleted relative to the IMF by stellar evolution, or the slope steepened by ejection of lower mass stars (see, e.g., \citealt{2012MNRAS.422.2246M} and \citealt{2012A&A...547A..23B}).
Here we will assume that the PDMF and the IMF are the same, but keep in mind these caveats. 

\begin{figure}
\epsscale{0.85}
\plotone{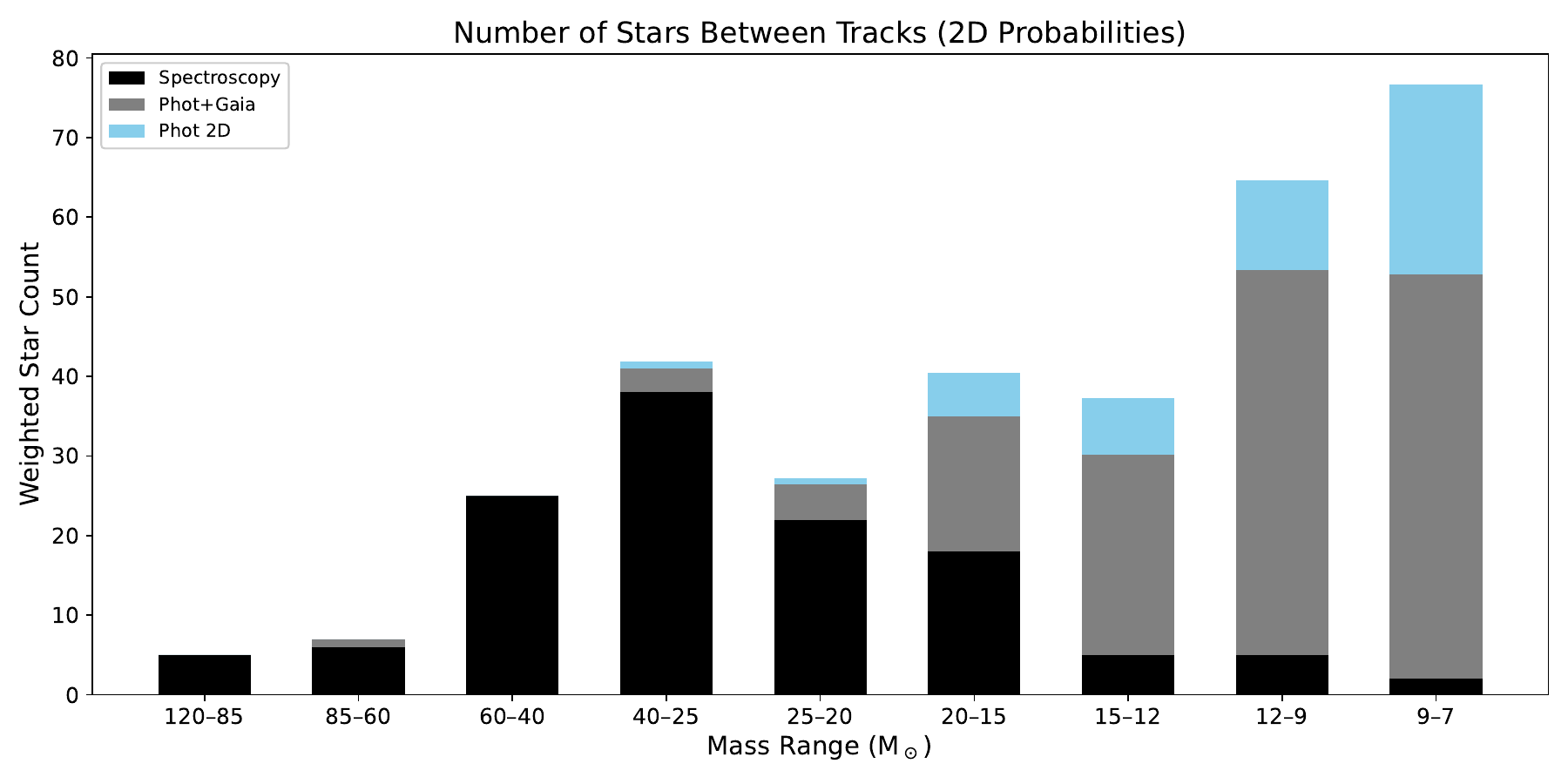}
\plotone{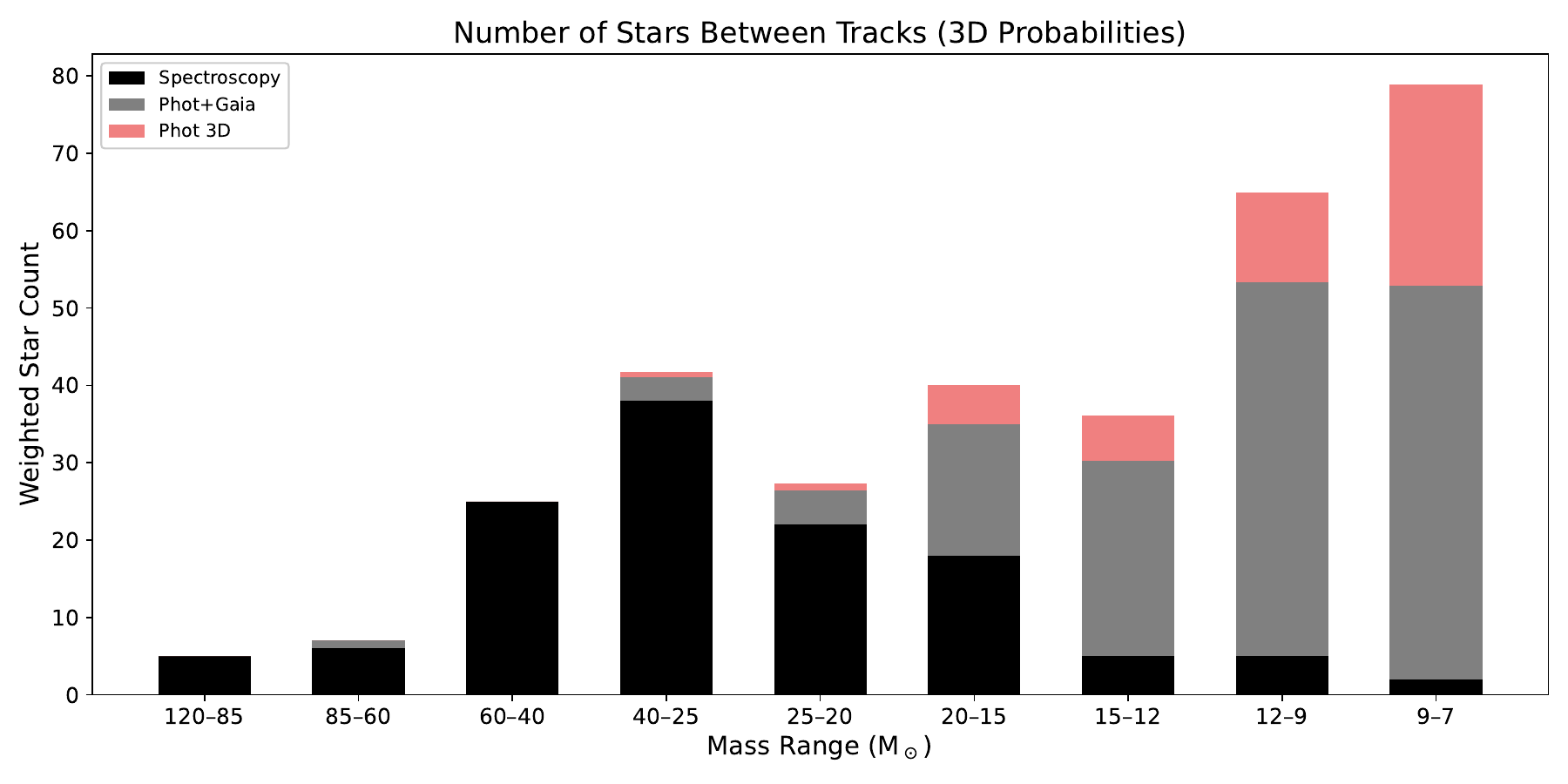}
\caption{\label{fig:hist} The contribution of spectroscopy and photometry to the number of stars in each mass bin.  The black bars denote the number of stars in each mass bin that were placed by means of spectroscopy, either by modeling or from the spectral types.  The grey boxes denote the stars without spectroscopy but with Gaia data indicating a membership probability.  The colored bars (blue for 2D and red for 3D) denote the contribution from stars whose membership probabilities were determined on the basis of location in the CMD.
}
\end{figure}

To derive the mass function slope of NGC~3603, we begin by counting the number of stars between successive massive tracks. We give these
numbers in Table~\ref{tab:Nums}, denoting which stars had been placed by spectroscopy, and which by
photometry.   The stars that were placed on the basis of spectroscopy are certain members given their early types,  but for the stars with only
photometry we count ``fractional stars" based upon their probability of membership, as determined in the previous sections.  In Figure~\ref{fig:hist} we show the number of stars in each mass bin that have been placed by spectroscopy and photometry. 

\begin{deluxetable}{l c c c c c c c}
\tablecaption{\label{tab:Nums} Number of Stars as a Function of Mass}
\tablehead{
\colhead{Mass ($M_\odot$)}
&\colhead{\# From Spectra}
&\multicolumn{3}{c}{\# From Photometry}
&
&\multicolumn{2}{c}{Total Number} \\ \cline{3-5} \cline{7-8}
& & 
\colhead{Gaia}
&\colhead{2D CMD}
&\colhead{3D CMD}
&
&\colhead{2D}
&\colhead{3D} 
}
\startdata
85-120   & 5  &  0   & 0    &  0   &&  5 &  5   \\
60-85    & 6  &  1   & 0    &  0   &&  7 &  7   \\
40-60    & 25 &  0   & 0    &  0   && 25 & 25   \\
25-40    & 38 &  3   & 0.9  &  0.7 &&  41.9 & 41.7 \\
20-25    & 22 &  4.4 & 0.8  &  0.9 && 27.2 & 27.3 \\
15-20    & 18 & 16.9 & 5.5  &  5.2 && 40.4 & 40.1 \\
12-15    &  5 & 25.2 & 7.1  &  5.9 && 37.2 & 36.1 \\
9-12     &  5 & 48.4 & 11.3 & 11.6 && 64.6 & 64.9 \\
7-9      &  2 & 50.9 & 23.8 & 26.0 && 76.7 & 78.9 \\
\enddata
\end{deluxetable}

Using the data from  Table~\ref{tab:Nums} we calculate the values of $\log{\xi}$, and show these in Table~\ref{tab:xi}.  We take the surface area to be 1.5$\times10^{-4}$ kpc$^2$,
corresponding to our survey area.  The uncertainties listed are simply the stochastically  $\sqrt{N}$ values to allow for the effects of small-number
statistics.  

\begin{deluxetable}{c c c c} [h!]
\tablecaption{\label{tab:xi} Mass Function}
\tablehead{
\colhead{Mass $m$}
&\colhead{$\overline{\log{m}}$}
&\multicolumn{2}{c}{$\xi(\log{m})$} \\ \cline{3-4} 
\colhead{($M_\odot$)}& \colhead{[$M_\odot$]} & \colhead{2D} & \colhead{3D} 
}
\startdata
85-120 &  2.004 &   5.35$^{+0.16}_{-0.26}$ &   5.35$^{+0.16}_{-0.26}$ \\
 60- 85 &  1.854 &   5.49$^{+0.14}_{-0.21}$ &   5.49$^{+0.14}_{-0.21}$ \\
 40-60 &  1.690 &   5.98$^{+0.08}_{-0.10}$ &   5.98$^{+0.08}_{-0.10}$ \\
 25-40 &  1.500 &   6.14$^{+0.06}_{-0.07}$ &   6.13$^{+0.06}_{-0.07}$ \\
 20-25 &  1.349 &   6.27$^{+0.08}_{-0.09}$ &   6.27$^{+0.08}_{-0.09}$ \\
 15-20 &  1.239 &   6.33$^{+0.06}_{-0.07}$ &   6.33$^{+0.06}_{-0.07}$ \\
 12-15 &  1.128 &   6.41$^{+0.07}_{-0.08}$ &   6.40$^{+0.07}_{-0.08}$ \\
  9-12 &  1.017 &   6.54$^{+0.05}_{-0.06}$ &   6.54$^{+0.05}_{-0.06}$ \\
  7- 9 &  0.900 &   6.67$^{+0.05}_{-0.05}$ &   6.68$^{+0.05}_{-0.05}$ \\
  \enddata
  \end{deluxetable}
  
\begin{figure}[b]
\plotone{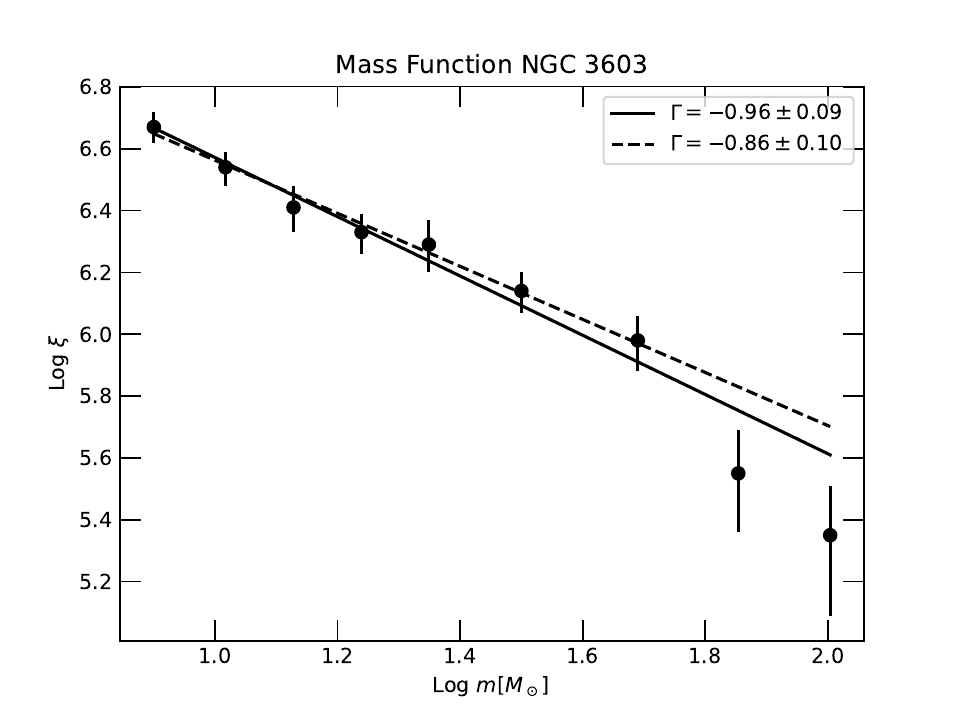}
\caption{\label{fig:IMF} The mass-function of NGC~3603.  The data from the 2D $\xi$ values are plotted; the 3D values are
indistinguishable.  The error bars denote the expected $\sqrt{N}$ stochastical variations. The solid line shows the linear fit for all
the data; the dashed line shows the fit ignoring the two highest mass bins.}
\end{figure}

In Figure~\ref{fig:IMF} we show the run of $\log{\xi}$ with the log of the mass.  We continue this down to the 7$M\odot$
track.\footnote{As mentioned earlier, our photometry is relatively complete to $V$=18.0-18.5 and $B$=19.0-19.5. With $E(B-V)$=1.2 to 1.8, we find that in
the worse case  the photometry extends down to an $M_V=-2.8$, or $\log L/L_\odot=3.3$.  As we see by comparison with Figure~\ref{fig:HRDspec}, this roughly corresponds to 7$M_\odot$.}  We see that values
through the 60$M_\odot$ track are well represented by a straight line.  The two highest mass bins,
60-85$M_\odot$ and 85-120$M_\odot$ are depleted relative to the other data. Fitting all the points,
weighting the data to take the stochastical uncertainties into account, yields a mass function slope
$\Gamma=-0.97\pm0.09$. Ignoring the two higher mass points results in a fit with $\Gamma=-0.87\pm0.10$.
The same result is obtained whether we use the 2D or 3D memberships
from the CMD for the stars without spectroscopy and without Gaia data. 

Furthermore, we see from Figure~\ref{fig:hist} that the stars whose placement is determined by photometry do not become dominant until
the 12-15$M_\odot$ bin ($\overline{\log{m}}=1.128$).  Yet there is no discontinuity present.  Indeed, if we had restricted our calculation only
to the three points (20-25$M_\odot$, 25-40$M_\odot$, 40-60$M_\odot$) we would get essentially the same slope, $-0.85$ rather than $-0.87.$

Throughout the 1990s and early 2000s the lead author and various collaborators measured the IMF slope in OB associations in the Magellanic Clouds and Milky Way.  Examples of well known regions include NGC~346 in the SMC \citep{1989AJ.....98.1305M}, 
R136 \citep{MH98}, NGC~6611 \citep{1993AJ....106.1906H},  Cyg OB2 \citep{1993AJ....105..980M} and h/Chi Per in the Milky Way \citep{2002ApJ...576..880S}.  The results of these studies are summarized in Table~1 of \citet{2011ASPC..440...29M}.  For the 25
OB associations and clusters with well defined IMFs, the values of $\Gamma$ ranged from $-0.7\pm0.2$ to $-2.1\pm0.6$. The
median value is $-1.3$, and the average weighted value is $-1.2$ with an RMS (1 $\sigma$ variation) of 0.3.  
No correlation was found
with metallicity or stellar density.  All 25 of the regions, as well as our study here, were done in the same manner, including simply
treating binaries as if they were single stars.   Of course, to derive the {\it actual} initial mass function these values must be corrected
for the influence of unresolved binaries, as emphasized by \citet{2007IAUS..241..109K}.  Lower mass stars may have been ejected, from the cluster core, steepening the IMF, but hopefully we have  avoided this by analyzing a region much larger than the central cluster.  So, while none of these are values completely right, they have all been carried out in an identical manner, and thus  comparisons are valid. 

Thus, with the result here we feel we can definitively answer the question posed in the title of the paper: no, the mass function of NGC~3603
is not top heavy.  The slope determined from stars in the range of 7-60$M_\odot$, $\Gamma=-0.9\pm0.1$, is no different than that of other well-studied regions analyzed in the same manner.

\section{Summary and Discussion}
\label{Sec-discuss}

Ground-based and ACS imaging provided photometry of 764 stars.  Most of these have Gaia astrometry that allows us to identify probability 
of membership in NGC~3603.  Using spectra collected over many years both at Magellan and HST, we have characterized the physical
properties of 129 of the brightest members, including 30 whose properties were determined by modeling.  For the remaining stars for which
there are no spectroscopy, we established membership probabilities either from Gaia data or from their location in the CMD.  The resulting
HRD shows  a very young (1-3~Myr) rich population of massive stars up to 120+$M_\odot$.  A few stars (e.g., Sh 25 and 
Sh 24) appear to be a few million years older, suggesting that the cluster formation was not strictly coeval.  This is similar to what we observe
in some other clusters, for instance NGC~6611 \citep{1993AJ....106.1906H}. We have analyzed the
mass function of this cluster using the same technique previously applied to 25 SMC, LMC, and Galactic OB associations (\citealt{2011ASPC..440...29M} and references therein) finding there is nothing unusual in its slope.

Rather than being top-heavy, if anything there are stars missing in the 60-120$M_\odot$ range.  Given the slope and intercept of our fit without these points, we calculate that there should be $25.9\pm13.7$ stars in that group.  We count only 11 (Table~\ref{tab:Nums}), 10 of which were placed by means of spectroscopy and one by photometry.  
In  the case of somewhat older clusters,
such as h and $\chi$ Per, the upper-most mass bin contains fewer stars than what we would expect based on extrapolation from other bins.  This is easily explained by stellar evolution depleting the higher mass bin.  However, here in NGC~3603 we would require something like
14 high mass stars to have gone through core-collapse.  While there is clearly a gas-free bubble surrounding the central cluster (Figure~\ref{fig:ebmv}),
we expect that 14 supernovae going off in the past few Myr in this region would have had a significantly more disruptive effect.  
While it is true that  some massive stars may undergo core-collapse without the usual fireworks, traditionally this was expected predominantly for the highest mass stars at sub-solar metallicity (e.g., \citealt{2003ApJ...591..288H}). In addition, recent work has suggested that the situation is more complex, with the detailed core structure and the time of core-collapse leading to non-monotonic  ``islands of explodability." For example, \citet{2016ApJ...821...38S} predict that stars with initial masses near 25 $M_\odot$ and 30 $M_\odot$ may undergo direct collapse, whereas several higher mass models may yield successful explosions (see their Figure 13). In addition, observational searches for failed SNe in nearby galaxies have thus far yielded only a few candidates which are predominantly YSGs or RSGs at intermediate ($\sim$25 $M_\odot$) masses (see, e.g., \citealt{2015MNRAS.450.3289G,2015MNRAS.453.2885R}). Regardless, fourteen is a lot of massive stars to have gone quietly.

This suggests that the formation of the highest high mass stars may actually have been slightly discouraged in NGC~3603.  This is not something we
see in the even richer region R136 on the LMC \citep{MH98,2020MNRAS.499.1918B}, 
but the metallicity of NGC~3603 is presumably 2$\times$ higher. What, then, could lead to
such a suppression? The question of what limits the ultimate mass of a star remains unanswered.  
\citet{1971A&A....13..190L} expected that 
accretion would be stopped by the effects of radiative heating, radiation pressure, and/or ionization.  They argued that the latter would be the dominant effect: as an H\,{\sc ii} region forms in the protostellar material, the temperature rises by several orders of magnitude,
and the resulting pressure stops the infall. Later, radiation pressure was thought to be the limiting
factor (e.g., \citealt{1987ApJ...319..850W}), but that too has been discounted: self-shielding is thought to allow the pressure to be
relieved through thin bubbles \citep{2009Sci...323..754K} or bipolar outflows \citep{2007ApJ...660..479B}.   
Recent papers suggest that radiatively-driven winds \citep{2018A&A...615A.119V} or UV line-driven disc ablation \citep{2019MNRAS.483.4893K,2024IAUS..361..550K} may provide the limiting factors.  Such mechanisms are metallicity dependent.
 Very high masses (200-300$M_\odot$) have been attributed to stars like R136a1 in the LMC based on atmospheric and evolutionary modeling (see, e.g., \citealt{2010MNRAS.408..731C, 2020MNRAS.499.1918B}).  However, even if this is correct, it may be that in the higher metallicity environment
 of the Milky Way (i.e., NGC~3603) that the limit may be lower, and may begin to depress star formation at the upper end of the IMF.  Why we do not see this same effect in other Galactic clusters may simply be the richness of NGC~3603: none of the previously analyzed regions are as strongly populated with such high mass stars.  Perhaps the ``upper mass limit" is not a hard cutoff, but rather affects the star formation probability down a few notches. 
 
 The ``observed" upper limit to the mass of a star is also poorly known.  There are many stars whose masses have been inferred to be well
 above 100$M_\odot$; these have become known as very massive stars (VMSs).   
 The most well known of these is the central star in the R136 cluster, R136a1, for which \citet{2010MNRAS.408..731C} derived a mass of 200-300$M_\odot$.  In fact, \citet{2025arXiv250615230K} suggest that the initial mass of R136a1 was $\sim 350M_\odot$, and that
 two other stars in the core, R136a2 and R136a3, had initial masses $>500M_\odot$!\footnote{Such numbers now approach the 1000$M_\odot$ mass once attributed to R136a \citep{1983ApJ...273..597S} before adaptive optics and HST showed it was a tight cluster.}  Based on evolutionary tracks and atmosphere modeling of their luminosities, \citet{2020MNRAS.499.1918B} identified 7 VMSs  in the R136 cluster.  In the Milky Way, the luminosities
 of such stars have sometimes been over-estimated due to undetected multiplicity. For instance, \citet{WalbornO2} estimate the mass of 
 Pismis 24-1 to be 200-300$M_\odot$, but the star was subsequently identified to be a previously unrecognized  triple system by \citet{2007ApJ...660.1480M}.  Still, given the preponderance of VMSs, and the claim of large binary fractions for massive stars (e.g., \citealt{sana12}),
 it is perhaps surprising that the highest masses reliably measured through binary motion remain in the 90-100$M_\odot$ range (see Table 5 in \citealt{MasseyA1}).  Finally, we recall that 
\citet{2012MNRAS.422..794E}  emphasized that binarity can affect any ``observed" upper mass limit, given that mass transfer can produce stars with
higher masses than could form by more pristine star formation processes.  

Regardless of why the two high mass bins are depressed relative to those of lower mass, our data suggest that the IMF of NGC~3603 is not top heavy as some have expected.  Whether there are actual variations in the slope of the IMF will doubtless remain a lively topic of debate for
the foreseeable future, but we note that both R136 and NGC~3603 provide examples of rather extreme environments where there is no
evidence of such variations.\footnote{A colleague of ours recently quipped that there is a 50~Mpc  ``rule" about the IMF slope: at any distance where it is currently impossible to verify the IMF via resolved stellar populations, studies will conclude that the IMF {\it must} vary.  Studies of regions
at distances small enough to actually do the measurements will always conclude that the IMF is universal. Of course, there are exceptions to this adage, most notably the recent study by \citet{2025ApJ...988..178L}.}

\begin{acknowledgements}

Lowell Observatory sits at the base of mountains sacred to tribes throughout the region. We honor their past, present, and future generations, who have lived here for millennia and will forever call this place home.   

The observations presented here were obtained over years at Las Campanas Observatory, and we are grateful to the excellent technical and logistical support we have always received there. We also acknowledge long-term support by both the Carnegie and Arizona Time Allocation Committees.  Obtaining the crucial IMACS multislit data was made possible with help from
Dr.\ Carlos Contreras, and the installation and use of the {\sc cosmos3} reduction software benefited from email exchanges with Drs.\ Gus Oemler and Daniel Kelson. 

Partial support for this work was provided by the National Science Foundation (NSF) through AST-2307594 awarded to P.M.  In addition, support for K.F.N. was provided from NASA through the NASA Hubble Fellowship grant HST-HF2-51516 awarded by the Space Telescope Science Institute, which is operated by the Association of Universities for Research in Astronomy, Inc., for NASA, under contract NAS5-26555.  Additional support for the analysis was provided by NASA through grants GO-6299.001 and AR-17553.001.
Support for the REU students was provided by the NSF through site awards 1004107, 1852478, and 1950901. 

We are grateful to many of our colleagues for help and advice throughout this work.  Dr.\ Joachim Puls provided us with his {\sc fastwind} code many years ago.  Our thoughts and discussion of the IMF benefited from conversations and correspondence
with Drs.\ Deidre Hunter, Gerhardt Meurer, Jacqueline Monkiewicz, Joel Parker, and Sumner Starrfield.   We are grateful to a quasi-anonymous referee for useful comments that improved the presentation.

Some assistance in Python programming (including the covariance probability assignment from Gaia data, the top panel of Figure~\ref{fig:ebmv}, and the random forest machine-learning classifier) was provided by ChatGPT-4 \citep{2023arXiv230308774O}.

This work has made use of data from the European Space Agency (ESA) mission
{\it Gaia} (\url{https://www.cosmos.esa.int/gaia}), processed by the {\it Gaia}
Data Processing and Analysis Consortium (DPAC,
\url{https://www.cosmos.esa.int/web/gaia/dpac/consortium}). Funding for the DPAC
has been provided by national institutions, in particular the institutions
participating in the {\it Gaia} Multilateral Agreement.

\end{acknowledgements}

\facilities{HST (STIS, ACS), Gaia, Magellan: Clay (MIKE, MagE), Magellan: Baade (MagE, IMACS), Swope (SITe No.\ 3 imaging camera, E2V CCD231-84 imaging camera), CTIO 1.0m (Y4KCam), CTIO 1.3m (ANDICAM)}

\bibliographystyle{aasjournalv7}
\bibliography{masterbib.bib}

\end{document}